\documentstyle[12pt]{article}
\input epsf.tex

\newcommand{\bmat}{\left(\begin{array}}
\newcommand{\emat}{\end{array}\right)}
\def\NPB#1#2#3{Nucl. Phys. B{#1} (19#2) #3}
\def\PLB#1#2#3{Phys. Lett. B{#1} (19#2) #3}

\def\PRD#1#2#3{Phys. Rev. D{#1} (19#2) #3}
\def\PRL#1#2#3{Phys. Rev. Lett. {#1} (19#2) #3}

\def\JHEP#1#2#3{JHEP  {#1} (19#2) #3}

\def\yzero{\smash{\hbox{$y\kern-4pt\raise1pt\hbox{${}^\circ$}$}}}

\def\beq{\begin{equation}}
\def\eeq{\end{equation}}
\def\beqa{\begin{eqnarray}}
\def\eeqa{\end{eqnarray}}

\def\-{\hphantom{-}}
\def\ov{\overline}
\def\s2{\frac{1}{\sqrt2}}

\def\beq{\begin{equation}}
\def\eeq{\end{equation}}
\def\beqa{\begin{eqnarray}}
\def\eeqa{\end{eqnarray}}

\def\Tr{{\rm Tr \,}}
\def\diag{{\rm diag \,}}
\def\IF{\relax{\rm I\kern-.18em F}}
\def\II{\relax{\rm I\kern-.18em I}}
\def\IP{\relax{\rm I\kern-.18em P}}
\def\IC{\relax\hbox{\kern.25em$\inbar\kern-.3em{\rm C}$}}
\def\IR{\relax{\rm I\kern-.18em R}}

\def\cp{{\cal P}}

\def\Dsl{\,\raise.15ex\hbox{/}\mkern-13.5mu D} 
\def\IZ{Z\kern-.4em  Z}

 \def\cp#1{\relax\ifmmode {\IP\kern-2pt{}_{#1}}\else $\IP\kern-2pt{}_{#1}$\=fi}

\catcode`\@=11
\newdimen\@rotdimen
\newbox\@rotbox

\def\@vspec#1{\special{ps:#1}}
\def\@rotstart#1{\@vspec{gsave currentpoint currentpoint translate
   #1 neg exch neg exch translate}}
\def\@rotfinish{\@vspec{currentpoint grestore moveto}}
\def\@rotr#1{\@rotdimen=\ht#1\advance\@rotdimen by\dp#1%
   \hbox to\@rotdimen{\hskip\ht#1\vbox to\wd#1{\@rotstart{90 rotate}%
   \box#1\vss}\hss}\@rotfinish}
\def\@rotl#1{\@rotdimen=\ht#1\advance\@rotdimen by\dp#1%
   \hbox to\@rotdimen{\vbox to\wd#1{\vskip\wd#1\@rotstart{270 rotate}%
   \box#1\vss}\hss}\@rotfinish}%
\def\@rotu#1{\@rotdimen=\ht#1\advance\@rotdimen by\dp#1%
   \hbox to\wd#1{\hskip\wd#1\vbox to\@rotdimen{\vskip\@rotdimen
   \@rotstart{-1 dup scale}\box#1\vss}\hss}\@rotfinish}%
\def\@rotf#1{\hbox to\wd#1{\hskip\wd#1\@rotstart{-1 1 scale}%
   \box#1\hss}\@rotfinish}%
\def\rotate{\@ifnextchar[{\@rotate}{\@rotate[l]}}
\def\@rotate[#1]#2{\setbox\@rotbox=\hbox{#2}\@nameuse{@rot#1}\@rotbox}

\catcode`\@=12

\begin{document}

\makeatletter \@addtoreset{equation}{section} \makeatother
\renewcommand{\theequation}{\thesection.\arabic{equation}}
\pagestyle{empty}
\rightline{FTUAM-2000/001; IFT-UAM/CSIC-00-02, DAMTP-2000-001}
\rightline{\tt hep-th/0001083}
\vspace{0.5cm}
\begin{center}
\LARGE{\bf
 A D-brane Alternative to the MSSM \\[10mm]}
\medskip
\large{G.~Aldazabal$^{1}$,
L.~E.~Ib\'a\~nez$^2$ and F. Quevedo$^3$
\\[2mm]}
\small{$^1$ Instituto Balseiro, CNEA, Centro At\'omico Bariloche,\\[-0.3em]
8400 S.C. de Bariloche, and CONICET, Argentina.\\[1mm]
$^2$ Departamento de F\'{\i}sica Te\'orica C-XI
and Instituto de F\'{\i}sica Te\'orica  C-XVI,\\[-0.3em]
Universidad Aut\'onoma de Madrid,
Cantoblanco, 28049 Madrid, Spain.\\[1mm]
$^3$ D.A.M.T.P., Wilberforce Road, Cambridge, CB3 0WA, England.
\\[3mm]}

\small{\bf Abstract} \\[3mm]
\end{center}

\begin{center}
\begin{minipage}[h]{15.0cm}
The success of  $SU(5)$-like  gauge coupling unification boundary
conditions $g_3^2=g_2^2=5/3 g_1^2$ has biased
most attempts to  embed the SM interactions into
a unified structure.
After  discussing the limitations of the orthodox approach,
we propose an alternative that appears to be quite naturally implied by
 recent developments based on D-brane
physics. In this new alternative: 1) The gauge
group, above a scale of order 1  TeV, is the minimal
left-right symmetric extension
$SU(3)\times SU(2)_L\times SU(2)_R\times U(1)_{B-L}$
of the SM; 2) Quarks, leptons and Higgs fields
come in three generations; 3) Couplings unify at
an intermediate string scale $M_s= 9\times  10^{11}$ GeV  with
boundary conditions $g_3^2=g_L^2=g_R^2=32/3\  g_{B-L}^2$. This corresponds to the natural embedding
of gauge interactions into D-branes and is different
from the standard $SO(10)$ embedding which
corresponds to $k_{B-L}=8/3$. Unification only
works in the case of three generations;
  4) Proton stability is automatic due to the presence
of $Z_2$ discrete R-parity and lepton parities.
A specific Type IIB string  orientifold model with the
above characteristics is constructed.
The existence of three generations is directly related to
the existence of three complex extra dimensions.
 In this model the
string scale can be identified with the intermediate
scale and SUSY is broken also at that scale due to the
presence of anti-branes in the vacuum.
 We discuss
a number of phenomenological issues in this model
including Yukawa couplings and a built-in axion
solution to the strong-CP problem.
The present framework could be tested by future accelerators
by finding the left-right symmetric extension of the SM at a scale of order 1 TeV.

\end{minipage}
\end{center}
\newpage
\setcounter{page}{1} \pagestyle{plain}
\renewcommand{\thefootnote}{\arabic{footnote}}
\setcounter{footnote}{0}

\section{Introduction}

The success of coupling unification extrapolations based
on the massless spectrum of the MSSM has greatly conditioned
the search for a realistic string vacuum. This search has been
shaped, to a great extent, by the fact that SM couplings seem to join at a scale
$M_X=2\times 10^{16}$ GeV, not far from the Planck mass $M_p$
 and by the necessity of identifying the string scale $M_s$
essentially with the Planck scale in heterotic model building.

In this situation, it appears natural to look for
perturbative heterotic vacua in which, below a scale
of order $M_X$, essentially only the MSSM remains.

 Recent
p-brane developments have
changed  our view of the possible ways to embed
SM physics into string theory. To start with, it has been realized
that the string/M-theory scale may be much below the Planck
and unification scales
\cite{witten,lykken,antoniadis,bajogut,sundrum,st,bachas,kakutye,benakli,biq,imr,dr}. This is because in the presence
of p-branes (like D-branes in Type II and Type I string theory)
gauge interactions can be localized in the world-volume of  D-branes
(e.g., a 3-brane), whereas gravitational interactions in general
live in the full ten (or eleven) dimensions. Then the largeness
of the Planck mass may be obtained even if $M_s<<M_p$
if there are  large compact dimensions.

Now, if $M_s<<M_X$, gauge coupling unification should
in principle take place  at the string scale $M_s$ and
 thus the nice unification of
MSSM couplings is lost.
Of course, this unification problem appearing for string
models if $M_s<<M_X$, could be taken  as an argument against them.
However, we think that  we should
first try to answer the following question:
Is there any simple alternative  framework
which is consistent with unification at a string scale
$M_s<<M_X$ ?
After all, the MSSM+big desert orthodoxy is not free of problems.
In fact, some unattractive features of the standard scenario are the following:

i) The quark/lepton generations come in three chiral copies whereas
the Higgs fields come only in one copy and are non-chiral.
The fact that there is a single Higgs set is crucial to
obtain correct unification predictions. This
asymmetry among quarks/leptons on one side and Higgs fields
on the other looks  quite {\it ad hoc}.

ii) The MSSM needs to be  supplemented by additional symmetries
like R-parity in order to ensure proton stability from
dimension four operators. It also needs additional symmetries
beyond R-parity to get stability against dimension five
operators.

iii)
To obtain a viable heterotic string unification, whose massless sector
is just the MSSM and includes the above symmetries, turns out to be
a very difficult task, if not impossible. All the models studied
up to now require a complicated study of possible scalar flat directions
and only very particular ones lead to something of that sort \cite{recent} .
The reason why dynamics should prefer such vacua with only one
set of Higgsses and  built-in discrete symmetries to suppress
too fast proton decay is unclear.

Given the above limitations of the orthodox approach, it seems
sensible to look for  (if possible, elegant) alternatives.
But to be really competitive with the MSSM scenario such
alternatives need to 1) improve some of the above problematic
aspects of the standard scenario and 2) have  a
nice and  consistent unification
of coupling constants. By the latter we mean that couplings unify at
the string scale without forcing the structure of the model by adding, for instance,
 unjustified extra mass scales or ad-hoc extra massless particles.

In the present paper we want to propose such an alternative to the
standard MSSM+ big desert scenario. We propose that above a scale
of order 1 TeV or so, the gauge group is that of the
minimal left-right symmetric extension of the supersymmetric standard
model, i.e.,$SU(3)\times SU(2)_L\times SU(2)_R\times U(1)_{B-L}$
\cite{mohapatra,mohap,bkmr}.
In addition all quarks, leptons and Higgs fields come in three generations.
Interestingly enough, this simple structure leads to
very precise unification of gauge coupling constants at an intermediate
scale of order $10^{12}$ GeV as long as the normalization
of the $ U(1)_{B-L}$ coupling is the one expected if such
gauge group is associated to a collection of D-branes. It is
important to remark that this normalization differs from the
one predicted by standard GUT left-right symmetric scenarios like
$SO(10)$.  

We also construct an explicit Type IIB orientifold
string compactification leading to the desired massless spectrum and
normalization of coupling constants.
In this model there is a simple explanation for the family
replication: there are three quark-lepton generations because there
are three complex compact dimensions and an underlying $Z_3$
orbifold.
These features  resemble the first
three-generation perturbative heterotic string models built, those
of ref.\cite{iknq}. Furthermore, this model has the interesting
property of having natural discrete symmetries including $R$-parity 
and $Z_2$ lepton numbers. Thus guaranteeing in a natural way the
stability of the proton, although allowing for other baryon number
violation 
processes, such as neutron-antineutron oscillations, sufficiantly
suppressed  to be consistent with current experimental bounds.
Finally, the structure of the model allows for candidate axion fields
with the right couplings to gauge fields needed to solve the strong CP
problem.

The structure of this paper is as follows. In chapter 2 we present
this alternative scenario which we call D-brane left-right symmetric
model. We also discuss the unification of coupling constants and show
how, if this model is correct,  new $Z$' and $W$' gauge bosons
corresponding to left-right symmetry should be found at future
or present colliders. In chapter 3 we present a particular
Type IIB orientifold model realizing the above scenario and study
the cancellation of $U(1)$ anomalies and generation of Fayet-Iliopoulos terms.
In this realization the unification scale is identified with a
string scale $M_s\propto 10^{12}$ GeV. The model has
also some anti-branes in the bulk which provide for hidden-sector
supersymmetry breaking at the same scale of order $M_s$.
We  study a number of phenomenological issues of this particular
orientifold model in chapter 4. This includes some aspects of the
structure of Yukawa couplings, $ SU(2)_R\times U(1)_{B-L}$ symmetry
breaking  and the presence of natural candidates for invisible
axions. In chapter  5 we present an  outlook and some final comments.

\section{The D-brane left-right symmetric model and coupling
unification}

As we discussed above, it has become recently clear that the string scale
could well be much below the Planck mass.
But, is there  any indication or advantage from a lowered string scale?
A particularly interesting alternative to the unification at
$M_X$ close to the Planck scale is getting unification  close to  the
geometric intermediate scale $M_I=\sqrt{M_WM_p}$. Indeed, if
the string scale is of order $M_s=\propto M_I$, gauge couplings should
unify at that scale.
 Now, as argued in ref.\cite{biq} , if there are
non supersymmetric brane configurations, the scale of supersymmetry breaking
would also be of order of $M_s=M_I$. This is interesting because
hidden sector supersymmetry breaking models also need to have
SUSY-breaking at the intermediate  scale. Thus
in this case the string, unification and SUSY-braking scales
would be one and the same.

 We would like to argue in what follows that the
intermediate scale idea is equally good than the standard one
in what concerns coupling unification, at least for
a  model  with the following structure:

i) The gauge group above  a  L-R symmetric scale
$M_R$ slightly above
the weak scale is the minimal left-right symmetric extension
of the SM: $SU(3)\times SU(2)_L\times SU(2)_R\times U(1)_{B-L}$.

ii) All quarks, leptons and Higgs fields come in three generations.
Thus the chiral multiplet content is three copies of
$(3,2,1,1/3)+({\bar 3},1,2,-1/3)$ $+(1,2,1,-1)+(1,1,2,+1)$ $+(1,2,2,0)$.

iii) The boundary conditions at the unification (i.e,.string)
scale are $g_3^2=g_L^2=g_R^2=32/3 g_{B-L}^2$.
This corresponds to a weak angle with $sin^2\theta(M_s)=3/14=0.215$.

In a model with the above characteristics one finds that gauge couplings
naturally unify at a scale of order the intermediate scale
$M_s\propto 10^{12}$ GeV as long as the left-right scale $M_R$ is
not far from the weak scale $M_W$.
 An important point to remark is that
the unification  boundary conditions are different from those
found in GUT schemes. Indeed in $SO(10)$-like schemes the
boundary conditions at unification are
 $g_3^2=g_L^2=g_R^2=8/3 g_{B-L}^2$ yielding the canonical
$sin^2\theta_W=3/8$.
A remarkable point we find is that {\it  the
new boundary conditions  we are proposing are precisely the ones which are
natural from the point of view of the embedding of the
gauge group in a D-brane scheme.}

Let us discuss in some more detail how coupling unification takes place.
The above mentioned boundary conditions $g_3^2=g_L^2=g_R^2=32/3 g_{B-L}^2$
have a simple group theoretical interpretation. They correspond
to the embedding of the left-right symmetric gauge group into
a non-semisimple structure:
\beq
 U(3)\times U(2)_L\times U(2)_R
\eeq
with unified coupling constants $g_3^2=g_L^2=g_R^2$ at some
mass scale (to be identified later on  with the string scale $M_s$ ).
This is in fact the structure one gets in models with gauge groups
living on D-branes, as we will discuss in the specific string
model below.
As we said,  the model contain three identical generations
under $U(3)\times U(2)_L\times U(2)_R$ with quantum numbers
$(3,{\bar 2},1)_{(1,-1,0)}
+({\bar 3},1,2)_{(-1,0,1)}$
 $+(1,{\bar 2},1)_{(0,-1,0)} +(1,1,2)_{(0,0,1)}$
$+(1,2,{\bar 2})_{(0,1,-1)}$, where the subindices
denote the charges with respect to the three $U(1)$'s.
We denote the $U(1)$ generators by $Q_3$, $Q_L$ and $Q_R$
respectively.
It is easy to check that two of them are anomalous
and only one of them, the linear combination
\beq
Q_{B-L}\ =\ -{2\over 3}\ Q_3 \ -\ Q_L \ -\ Q_R
\label{b-l}
\eeq
is anomaly free
\footnote{In string theory the other  two (anomalous) $U(1)$'s
become massive and decouple due to a generalized Green-Schwarz
mechanism. See the discussion in chapter 3.}
. This is just the familiar $(B-L)$ of
left-right symmetric models which is related to weak hypercharge
by $Y=-T_R^3+Q_{B-L}/2$.
Now, notice that, if we normalize the
original $U(n)$ generators
 in the fundamental representation
 $T_a$ to $TrT_a^2=1$, the normalization of the $U(1)$'s
are $TrQ_3^2=3$, $TrQ_L^2=TrQ_R^2=2$. Then, the
normalization of $U(1)_{B-L}$ compared to that of the
non-Abelian generators is $k_{B-L}=2TrQ_{B-L}^2=32/3$,
as remarked above. Notice this implies a
hypercharge normalization $k_1=k_R+1/4k_{B-L}=11/3$, and hence
a tree level weak angle $sin^2\theta _W= 3/14=0.214$.

Let us study now the one-loop corrections to the couplings.
In between the scales $M_R$ and $M_s$ the gauge group is
$SU(3)\times SU(2)_L\times SU(2)_R\times U(1)_{B-L}$ and the
above chiral field content gives rise to the following
one-loop $\beta$-function coefficients $B_a$ :
\beq
B_3\ =\ -3 \ ;\ B_L\ =\ +3 \ ;\ B_R\ =\ +3 \ ;\  B_{B-L}\ =\ +16
\label{br}
\eeq
In between the weak scale $M_W$ and  $M_R$ the gauge group will be that of
the SM with $\beta $-function
coefficients $b_i$. Then the one loop running yields:
\beqa
& \sin^2\theta _W(M_Z)\ =&\ {1\over {1+k_1}}(1\ +\ k_1{{\alpha _e
(M_Z)}\over {2\pi }}\ [(B_L- \frac{1}k_1B_1')\ \log({{M_s}\over
{M_R}}) \ \nonumber \\ & &  +\ (b_2-\frac{1}k_1b_1)\ \log({{M_R}\over
{M_Z}}) \ ]
   \\
&{ 1\over {\alpha _e(M_Z)} }\ -\
{{1+k_1}\over {\alpha _3(M_Z)}}\ =&\
 {1\over {2\pi }}\ [
(b_1+b_2- (1+k_1)b_3)\ \log({{M_R}\over {M_Z}}) \ \nonumber \\
& & +\
(B_1'+B_L-(1+k_1)B_3)\ \log({{M_s}\over {M_R}})\ ]
\label{senos}
\eeqa
where one defines
\beq
B_1'\ =\ B_R\ + \ {1\over 4} B_{B-L}
\eeq
and $k_1=k_R+1/4\ k_{B-L}$.  With the minimal particle content described
above one has $B_1'=7$.
Let us suppose for the moment that, below the $M_R$ scale
down to $M_Z$, we were
left just with the content of the MSSM.
We would then have (for $k_1=11/3$, corresponding to $k_{B-L}=32/3$):
\beqa
& \sin^2\theta _W(M_Z)\ =&\ {3\over {14}}(1\ +\ {{\alpha _e
(M_Z)}\over {2\pi }}\ [4\ \log({{M_s}\over
{M_R}}) \   - \  {{22}\over 3 }\ \log({{M_R}\over
{M_Z}}) \ ])
   \\
&{ 1\over {\alpha _e(M_Z)} }\ -\
{{14}\over {3\alpha _3(M_Z)}}\ =&\
 {1\over {2\pi }}\ [
26 \ \log({{M_R}\over {M_Z}}) \
 +\
24 \ \log({{M_s}\over {M_R}})\ ]
\label{senos113}
\eeqa
Now, using as input  $\alpha _e(M_Z)^{-1}=127.934\pm 0.027$  in ref.\cite{PDG}
(in particular its central value), one can plot the predicted
$sin^2\theta_W(M_Z)$ versus
$\alpha_3(M_Z)$ and compare it to the experimental data
for those two quantities. This is done for several values of $k_{B-L}$ in fig.1,
where the data plotted correspond to the world-average in ref.
\cite{PDG} with two standard deviation errors.
It may be observed that, for the particular normalization corresponding
to an embedding of the left-right symmetric interactions into
D-branes ($k_{B-L}=32/3$), a very nice agreement
with the data is found. Very slight departures from this value are
ruled out. The  unification scale corresponding to the
successful results is of order $M_s=9\times 10^{11}$ GeV.
 We show for comparison a similar plot for the MSSM
obtained from the one-loop formulae in fig.2. Again, for the
MSSM standard normalization $k_1=5/3$ the prediction nicely goes
through the data points, in this case for a unification mass
of order $M_X=2\times 10^{16}$ GeV.
Thus we may conclude that, within the approximations made, the
D-brane left-right symmetric model is remarkably successful
in obtaining appropriate gauge coupling unification.

\begin{figure}
\centering
\epsfysize=10cm
\leavevmode

\epsfbox{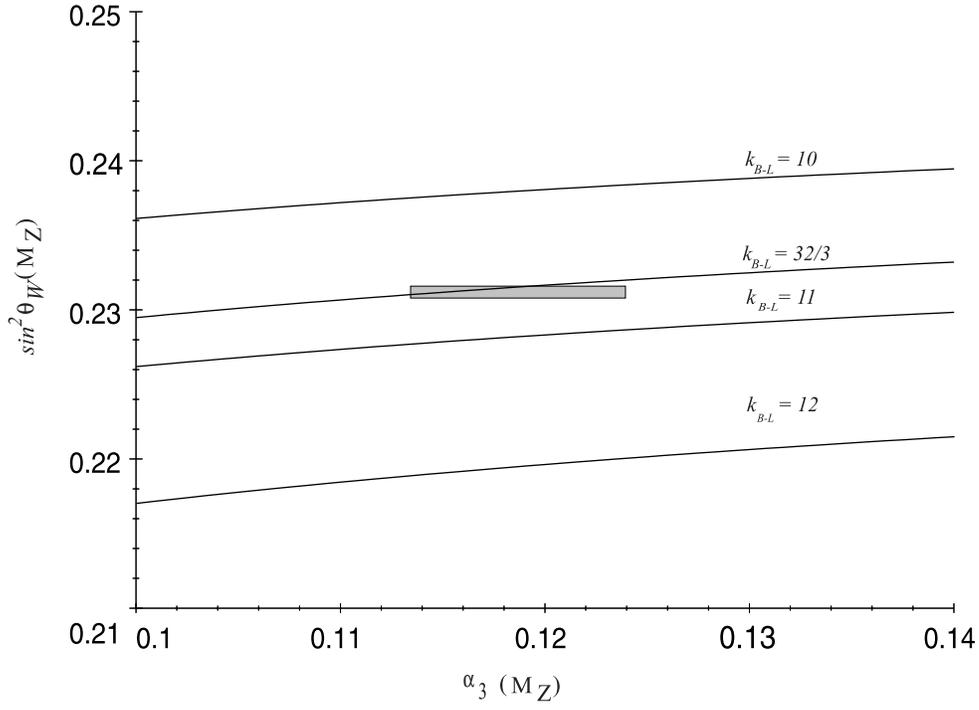}
\caption[]{$sin^2\theta_W (M_Z)\ vs. \alpha _3 (M_Z)$ for different values of
$k_{B-L}$.}
\end{figure}

\begin{figure}
\centering
\epsfysize=10cm
\leavevmode

\epsfbox{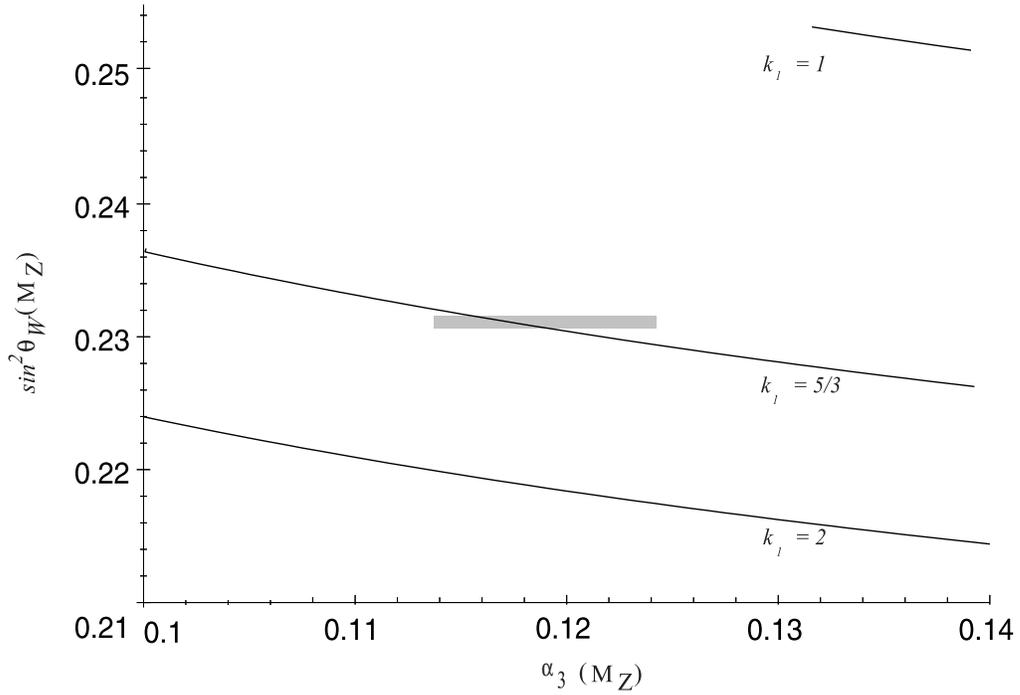}
\caption[]{$sin^2\theta_W (M_Z)\ vs. \alpha _3 (M_Z)$ for different values of
$k_{1}$ for the MSSM.}
\end{figure}

It  can also be  checked that, within this scheme, both the SUSY-breaking
scales and the left-right scale $M_R$ cannot be much above the
1 TeV scale. Indeed, let us now use also as input the world average central value for the
weak angle,  $sin^2\theta_W(M_Z)=0.23117\pm 0.00016$ as well as
$\alpha _e(M_Z)$ and let us plot the predicted
$\alpha _3(M_Z)$ as a function of the left-right scale $M_R$
for various values of the SUSY-breaking mass $M_{sb}$.
 This is shown in figs. 3 and 4, where we have assumed that below the
$M_R$ (and above the $M_{sb}$) scale one is left with the
particle content of the MSSM. In fig. 3 we consider a universal SUSY-breaking
 threshold $M_{sb}$  below which the non-SUSY  SM
is obtained. Fig. 4 plots the same quantities but now with two
SUSY-breaking thresholds one ($M_{sb1}$) for the coloured
SUSY-particles (squarks and gluinos) and a second lower one $M_{sb2}$
for non-coloured ones.
 We  observe that for
 large values of $M_R$, a low value for the SUSY threshold
seems to be required to get a consistent value for $\alpha_3(M_Z)$.
Thus either the SUSY threshold or the
left-right symmetric threshold (or both) should be below the
1 TeV scale.
Although we have only used one-loop formulae and have taken
step functions for the different thresholds we expect that those
refinements will not substantially change the main conclusion
that couplings nicely unify as long as the $M_R$ scale is not much above 1
TeV.

Several comments are in order:

i)
Notice  that we have not played around with the addition of extra
mass scales and /or extra particles beyond the three quark/lepton/Higgs
generations in order to get satisfactory unification.
  So this is not a mere adjustment of the model to get nice coupling
unification, it appears naturally as long as $M_R$ is not much higher than
1 TeV.

ii) In order to obtain the above interesting unification results it is
crucial to use the boundary conditions
 $g_3^2=g_L^2=g_R^2=32/3 g_{B-L}^2$.
 Thus if we  would have used the standard
GUT conditions, with $k_{B-L}=8/3$ instead of $32/3$, we would  obtain
 $M_s\propto 10^{13}$ GeV and $sin^2\theta _W(M_Z)=0.34$
(for input $\alpha _3(M_Z)=0.119$ and $M_R\propto 1$ TeV).
As we have mentioned, the boundary conditions which work
correspond to those expected when the gauge group is embedded
inside a collection of Type IIB D-branes.

iii) The tree level result $sin^2\theta_W=3/14=0.214$ is quite close already
to
the experimental result 0.231. One loop effects should then be
{\it small and positive} (i.e., of order 8\% ). This contrasts with  the
standard MSSM/GUT case, where  the tree level result $sin^2\theta_W=3/8=0.375$
departs from the experimental number. In this case the loop corrections
must be {\it large and negative} (of order 62\% ). In this
connection notice that the $sin^2\theta _W$ loop corrections have
two pieces in our case, one positive and proportional to
$log(M_s/M_R)$ and one negative and proportional to
$log(M_R/M_Z)$. Since a positive correction is needed in order to get
agreement
for $sin^2\theta _W$, this is the hidden reason why gauge coupling unification
requires in our case large $log(M_s/M_R)$ but small $log(M_R/M_Z)$.

iv) Coupling constant unification works only for three
generations of quarks, leptons and Higgs fields. Thus there is a connection
between generation number and unification
(unlike the MSSM which is insensitive, at one loop, to the
number of quark-lepton generations). Let us  denote
the number of quark/lepton/Higgs generations as $n_g$. Thus, $\beta$-function
coefficients become  now $B_3=-9 +2n_g$, $B_L=B_R=-6+3n_g$,
$B_{B-L}=(16n_g)/3$ and $B_1'= -6+(13/3)n_g$. The combinations relevant
for the running are
\beq
[B_L\ -\ {3\over {11}}\ B_1'] \ =\ {4\over {11} } \ (5n_g\ -\ 12)
\ ;\  [B_1'\ +\ B_L\ -{{14}\over 3}\ B_3 ]\ =\ 30\ -\ 2n_g
\eeq
Now, notice that in order to have  a positive correction for
$sin^2\theta_W$ as required, we need  $n_g\geq 3$. On the
other hand, for $n_g\geq 4$ either $sin^2\theta_W$ becomes  too large
or $log(M_s/M_R)$ becomes too small to be compatible with unification.

v) We have assumed that at a scale $M_R$ the gauge group is broken to
that of the SM but we have not specified what the fields, giving rise to such breaking, are.
As we will show in the specific
string construction below, there are simple additions to the model
(e.g., from the presence of some D-branes
in the bulk, which lead to  non-chiral particle content)
which contain the required fields for this breaking without
 modifying the runnings at one loop, thus preserving the
interesting results described above.

The above new scheme may probably be obtained in different classes
of string models involving D-branes with a gauge group
$U(3)\times U(2)\times U(2)$.
One of the most  interesting points we find is that
 to  construct a specific  four-dimensional
Type I string model with precisely that massless spectrum and with the
necessary gauge coupling boundary conditions is very simple. In fact,  such a
model was briefly discussed in section (4.1) of ref.\cite{aiq}.
We  do not claim however that this is the only possible realization
of the D-brane left-right symmetric scenario here introduced.
Nevertheless, we think it is worth studying such a model, since it may provide clues of more general features of the
scheme.
We now describe the construction of the mentioned Type IIB orientifold
realization.

\begin{figure}
\centering
\epsfysize=10cm
\leavevmode

\epsfbox{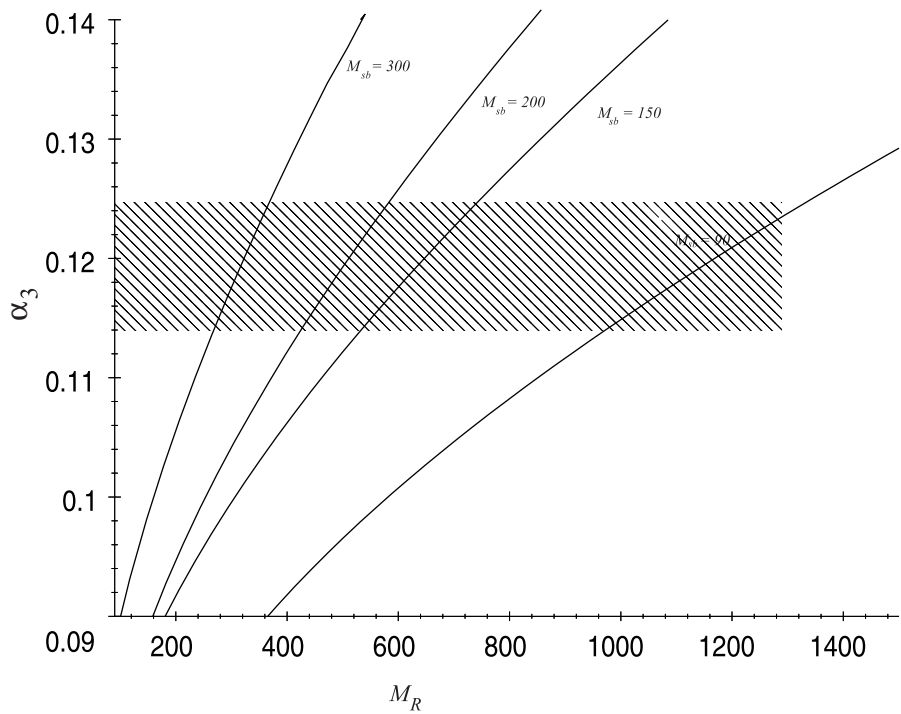}
\caption[]{$\alpha_3(M_Z)  vs. M_R$ for different values of
a  universal SUSY-threshold $M_{sb}$. Here LR symmetry is broken above
supersymmetry }
\end{figure}

\begin{figure}
\centering
\epsfysize=10cm
\leavevmode

\epsfbox{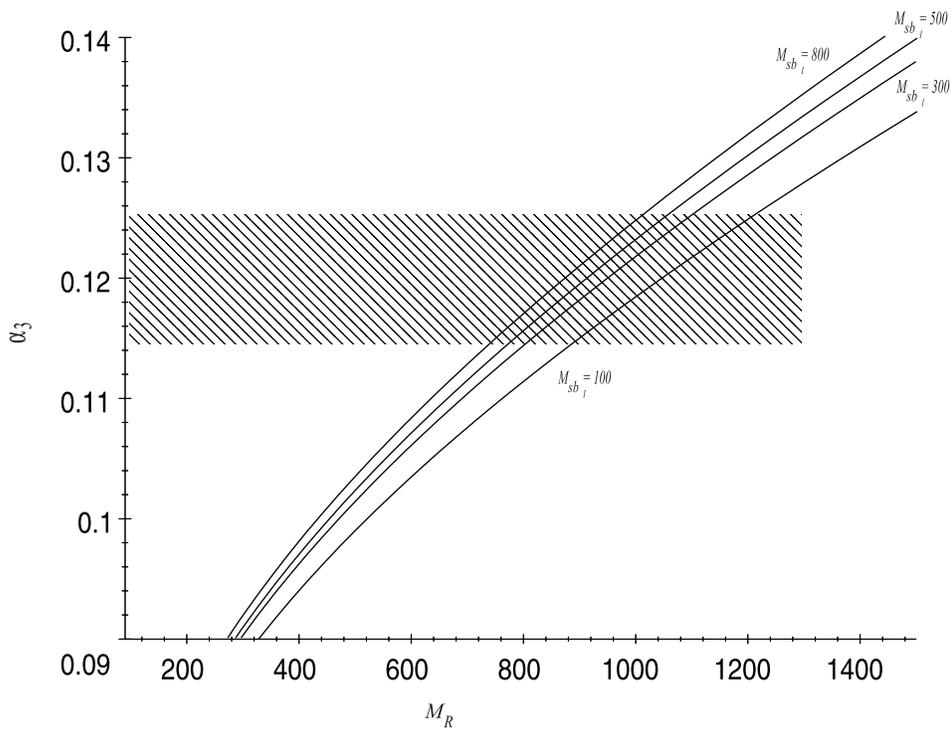}
\caption[]{$\alpha_3(M_Z)  vs. M_R$ for a non-colored SUSY particle threshold
 $M_{sb2}=100$ and different values of a  coloured 
 SUSY
threshold $M_{sb1}$. }
\end{figure}

\section{A Type IIB orientifold with left-right symmetry}

\subsection{The LR orientifold model}

The model we are interested in is a $Z_3$ Type IIB orientifold with both D-branes and anti
-D-branes. A general discussion of such kind of models is given in Ref.
\cite{au,aiq}
where we refer the reader for notation and details of the construction
\footnote{For other constructions involving anti-branes
see \cite{ads} }. Here we only present
a brief description in order to settle the general framework.

A $Z_3$ Type IIB orbifold, in four dimensions \cite{ang},
 is obtained by dividing closed Type IIB string theory
compactified on a six dimensional torus $T^6$, by the discrete symmetry group $Z_3$.
The  orientifold model is obtained  by further dividing the orbifoldized
string
by world sheet orientation reversal symmetry \cite{orient}.
The twist eigenvalues, associated to complex coordinates
$ Y_a \ a=0,1,2$ are chosen as $v=\frac{1}3(1,1,-2)$ in
order to leave $N=1$ supersymmetry in four dimensions.

The general picture is that the above procedure
 leads to a Klein-Bottle unoriented world sheet. Amplitudes
computed on such a surface contain unphysical
 tadpole like  divergences which can be interpreted as unbalanced charges
carried by RR form potentials. Thus, in order to
cancel such divergences, D9-branes, carrying opposite charges must be
introduced.  Moreover, D5-branes and anti-D5-branes can be consistently included.
Even if they are not required (in this $Z_3$ case)   for tadpole
cancellation, they open the way for achieving interesting
supersymmetry breaking patterns and at the same time provide
new possibilities for model building.

Let us be more explicit. An open string state is denoted  by $|\Psi, ab \rangle \lambda ^{pq}
_{ab}$ where  $\Psi$ refers to
world-sheet degrees of freedom whereas $a,b$ are Chan-Paton
indices associated to  the open string endpoints lying on
D$p$-branes and D$q$-branes respectively
\cite{ang,IIBorient,afiv} . $\lambda ^{pq}$ is the
Chan-Paton, hermitian matrix, containing the gauge group structure
information. Analogously, $\lambda ^{\ov p \ov q}$ ($\lambda ^{\ov
p q}$) is introduced for open strings ending at D$\ov p$, D$\ov
q$-antibranes ( D$\ov p$ antibrane,  D$ q$-brane, etc.).

The $Z_3$  action (denoted by $\theta$)  that twists the internal complex coordinates has a corresponding action on
 Chan Paton matrices represented
by unitary matrix  $\gamma _{\theta ,p}$, namely
$\theta: \lambda ^{pq} \rightarrow \gamma _{\theta,p} \lambda ^{pq} \gamma^{-1}_{\theta,q}$.
Moreover, Wilson lines, wrapping along internal tori directions
can also be included and also have a matrix representation when
acting on Chan-Paton factors.

Consistency under group algebra operations and the requirement of
cancellation of RR tadpoles leads to constraints  on the possible
twist matrices. Tadpole cancellation, in the $Z_3$ case we are
discussing,
imposes  the number of nine branes to be  32 and the requirement that the number of D5-branes and
anti-D5-branes must be equal. Moreover, cancellation of twisted
tadpoles requires
\beq
\Tr ({\cal W})^a\gamma_{\theta,9}
 +
3(\Tr\gamma_{\theta,5, {a,i}}- \Tr\gamma_{\theta,\bar{5},{a,i}}) = -4
\label{tadpzz} \eeq
for $a=0,1,2$.

Here we have allowed for the possibility of having a Wilson line,
represented by the matrix ${\cal W}$ on Chan -Paton matrices,
wrapping along the direction $e_1$.
We denote with  ${a,i}$ with $a, i=0,1,2$ the nine orbifold
fixed points  in the first and second complex planes.
For a given $a$, $ i=0,1,2$ label the subset of fixed points
that feels the twist $({\cal W})^a\gamma_{\theta,9}$.

Generic solutions to these equations are discussed in \cite{aiq}.
Here we consider the specific model characterized by
$\gamma_{\theta,9}=({\tilde \gamma_{\theta,9}}, {\tilde
\gamma_{\theta,9}}^*)$, and
${\cal W}= (\tilde {\cal W},{\tilde {\cal W}}^*)$ where $ *$ denotes the complex
conjugate and

\beqa\label{g9wl}
{\tilde \gamma_{\theta,9}}& = & \diag ( {\alpha I_3,{\alpha }^2 I_2,I_2 ,I_2,
 \alpha I_7})\\
{\tilde {\cal W}}&=&  \diag ({I_3,I_2,I_2,I_2,I_7})
\eeqa
Also,
\begin{equation}\label{g5fp}
\gamma_{\theta,5, {2,i}}=  \diag ( \alpha ,{\alpha}^2 )
\end{equation}

where $\alpha = e^{2i\pi /3}$.

It can be easily checked that such a choice satisfies the tadpole
cancellation constraint \ref{tadpzz}. Notice that there are two
5-branes ($\Tr\gamma_{\theta,{5},{2,i}}=-1$)  stuck at each of the
three fixed points of the type $(2,i) \equiv (-1,i)$.
The effective twist is thus ${\cal W}^{-1} \gamma_{\theta,9}$.
There are no 5-branes at the other six fixed points
($ \Tr (\gamma_{\theta,9})=\Tr ({\cal W})\gamma_{\theta,9}=-4$).
Since we must have the same number of branes and antibranes, six
anti-5-branes must be present in the bulk.

It is always possible to add an extra Wilson line in the third
complex plane in such a way that anti-brane sector is gauge
decoupled from the 9, 5-branes sectors. In this way,  anti-branes in
the bulk, which lead to a non-supersymmetric spectrum, provide a``hidden sector" which will transmit supersymmetry breaking through
gravitational interactions to the ``observable" brane sector.
 An alternative description of such a decoupling situation
 can be achieved by performing a
 T-duality transformation in the third complex dimension. Thus,
 9-branes become 7-branes located at the origin ($Y_3=0$) in the third complex
 plane  with their world volume including
 the first two complex planes. 5-branes (antibranes)  turn into 3-branes
 (antibranes) which can be located anywhere on the compact space.
This is pictorically described in figure 5 \footnote{Notice that in this figure
extra
`bulk' branes are added (see below) and only two antibranes are shown,
the other antibranes are located at the image of these two under the
combination
of the $Z_3$ defining the orbifold and the orientifold $\Omega$
(or $(-1)^{F_L}\Omega R_3$ in the 7,3-brane picture) which
is a further $Z_2$ symmetry, making for a total of $2\times 3\times 2=12$
antibranes
which equals the number of 3-branes.}.

\bigskip
\begin{figure}
\centering
\epsfysize= 9 cm
\leavevmode

\epsfbox{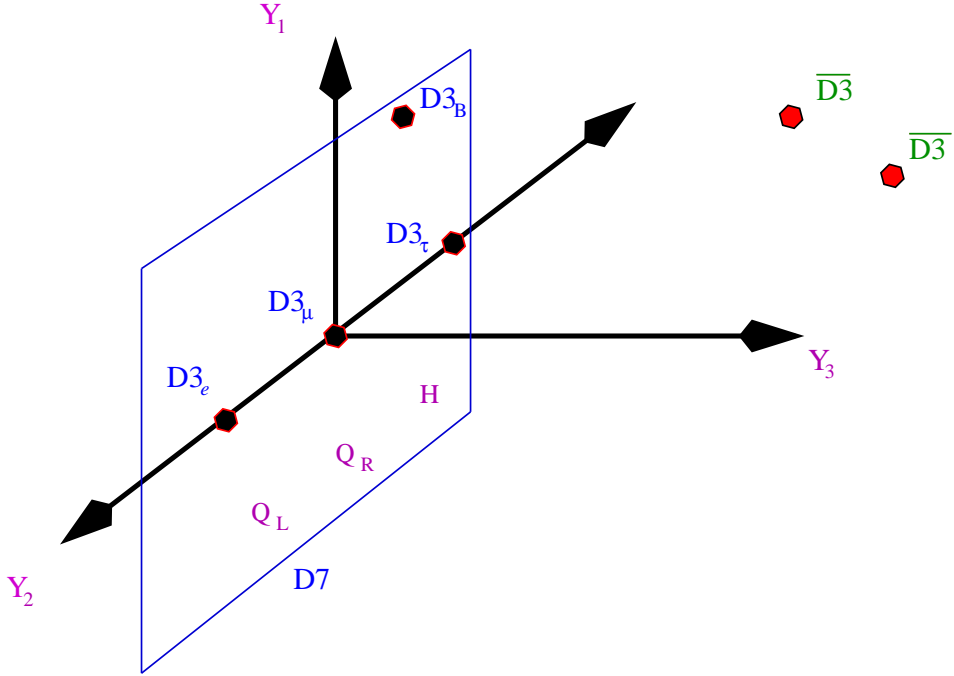}
\caption{D-brane configuration of the model discussed in
the text. The gauge group, Quarks and Higgs fields live
in the worldvolume  for the D7-brane
 whereas leptons are located at the three fixed points
 in the complex $Y_2$ dimension. Bulk D$3_{bulk}$ branes, leading
to $SU(2)_R\times U(1)_{B-L}$ breaking also live
in the $Y_1,Y_2$ hyperplane whereas anti-3-branes
live away in the bulk of the third ($Y_3)$ complex dimension.}
\end{figure}

 Hence, if  bulky anti-3-branes are placed away from $Y_3=0$ their worldvolume will not overlap
that of the  supersymmetric 7,3-brane sectors. Thus bulky,  non-supersymmetric fields, will decouple
from the supersymmetric sector and will only couple to it through
the exchange of closed string interactions.  This
 is the T-dual description of the Wilson line accounted for in the
 preceding paragraph.
 We adopt this point of view in what  follows. Twist matrices
in the T-dual description are  obtained just by exchanging $7 \leftrightarrow 9$,
 $3({\bar 3}) \leftrightarrow 5 ({\bar 5})$ indices.
 Thus, spectrum and interactions  are the same in both descriptions.

Twist matrices can be described in terms of associated shift vectors
(\cite{afiv, aiq}). Such a description, which  naturally appears when
a Cartan-Weyl basis is chosen for the group algebra, is especially adapted
for computing the spectrum \cite {afiv,aiq}. Thus, twist matrices above (now for
 7,3-branes) correspond to

\beqa
V_7\ =\  \frac{1}{3}\, (1,1,1; -1,-1; 0,0; 0,0; 1,1,1,1,1,1,1)\\
W\ =\  \frac{1}{3}\, (1,1,1; 1,1; 1,1; 0,0; 0,0,0,0,0,0,0)
\eeqa
and simply $V_{3,(2,i)}= \frac{1}{3}$ for matrices in \ref{g5fp}.

We then have for $ ({\cal W})\gamma_{\theta,9}$ and
${\cal W}^{-1}\gamma_{\theta,9}$
\beqa
V_7+W\ =\ \frac{1}{3}\, (2,2,2; 0,0; 1,1; 0,0; 1,1,1,1,1,1,1)\\
V_7-W\ =\ \frac{1}{3}\, (0,0,0; 1,1; 2,2; 0,0; 1,1,1,1,1,1,1)
\eeqa
respectively.
The total observable gauge group is $G=G_7\times G_3  $
with the {\bf 77}-sector brane group being
$G_7=U(3)\times U(2)_L\times U(2)_R\times SO(4)\times U(7)$,
while the ${\bf 33}_{2,i}$ sectors, with branes trapped at $(2,i)$
($i=0,1,2$)
fixed points leads to  $G_3=U(1)_{2,0}\times U(1)_{2,1}\times U(1)_{2,2}$.
The six anti-3-branes in the bulk give rise to a single $Sp(2)$: two of them give rise to
a $Sp(2)$ and the other four are just $Z_3$ mirrors of the first  ones.

\vspace*{1cm}

\begin{table}[htb]%
\scriptsize
\renewcommand{\arraystretch}{1.2}
\begin{center}
\begin{tabular}{|c|c|c|}
\hline
{\bf 77} sector
& {\bf 37} sector  & {\bf 33} sector   \\
\hline\hline & & \\
 $3[(3,2,1)+(\bar 3,1,2) +(1,2,2) $ & $
3[(3,1,1)+(\bar 3,1,1)+ (1,2,1)+(1,1,2)$
 & $3[(1)]$ \\
$+(1,21)'+(4,\bar 7)']$ & $+(1,7)'+(4,1)']$ &  \\
\hline\hline & & \\
{\bf 37} sector (bulk) & {\bf 33} sector (bulk) & ${\bf \bar 3\bar 3}$ sector \\ \hline\hline
& & \\
 $(3,1,1;2)+(1,2,1;2)+(1,1,2;2)$
& $2(1)$ & $f_-: (1)\quad f_+: 2(3)+(1)$\\  $+
 (1,7;2)'+h.c.+(4,1;2)'$ & $ + (3)$ & $s: 2(1)+(3)$\\ \hline
 \end{tabular}
\end{center} \caption{Spectrum of Left-Right model. We present the
quantum numbers  under the groups
$SU(3)_c\times SU(2)_L\times SU(2)_R \times SO(4)'\times SU(7)'$ on the
7-branes,
$U(1)^3$ on the trapped 3-branes, $Sp(2)$ on the bulk 3-branes and
$Sp(2)^2$ from the $\bar 3$-branes.
\label{tabps1} }
\end{table}

Let us analyze  the  supersymmetric part of the spectrum  (see \cite{aiq}).
We  find the chiral content

\beqa
& &{\bf 77} sector    :\\
& & 3[(3,2,1,1/3\ )+(\bar 3,1,2,-1/3\ ) +(1,2,2,0\ )]+
3[(1,21 \ )'+(4,\bar 7\ )']\nonumber
\eeqa
where we have indicated the representations under the Left-Right
group $G_{LR}=SU(3)\times SU(2)_L\times SU(2)_R\times U(1)_{B-L}$ and
$G'= SO(4)'\times U(7)'$. The Abelian factor $U(1)_{B-L}$
is generated by $  Q_{B-L}= - \frac{2} 3 Q_3-Q_L-Q_R$ introduced
in \ref{b-l}
  where $Q_3,Q_L,Q_R$ are the
 generators of the Abelian factor in the corresponding unitary
 groups. As mentioned,  $Q_{B-L}$ is identified with $B-L$ symmetry
generator and it can be shown (see next section)
 to be non anomalous.
The  Standard Model hypercharge $Y$ and electromagnetic charge are
 thus obtained as
\beqa\label{hych}
  Y & = & \frac{Q_{B-L}}2-T^3_R \\
  Q_{em} & =&  Y +T^3_L \noindent
\eeqa
 where  $T^3_{R,L}$ are the diagonal generators of $SU(2)_{R,L}$.

 We observe that the  ${\bf 77}$ sector contains the standard three quark
 generations plus a set of three chiral Higgs fields $(1,2,2,0)$.
 The factor three here is associated to the three compact complex
 dimensions $Y_a$ $(a=0,1,2)$ each one of them feeling the same
 orbifold twist.
\beqa
& &{\bf 37}_{2,i} sector: \\
& & (3,1,1,-2/3)_{-1}+(\bar 3,1,1,2/3)_{-1}+
(1,2,1,-1)_1+(1,1,2,+1)_1
+[(1,7)_1'+(4,1)_1']\nonumber
\eeqa
 A subindex indicates the charge with respect to the
$U(1)_{3,(2,i)}$ 3-brane group. Since $i=0,1,2$, we will have three
identical copies. Hence, these sectors provide three generations of
standard leptons.

We display the summary of the massless spectrum of the model
in Table 1 \footnote{ In Tables 1,2,3  we also display the extra
massless fields which might appear
if, in addition to the branes discussed above, there are further
3-branes living in the bulk (see fig. 5)  of the first two complex dimensions
(but at $Y_3=0$). These 3-branes may be used to break the left-right
symmetry down to the SM, as we discuss in  next chapter.
They do not modify one-loop coupling unification, though.}.
Each of the three ${\bf 33}$ sectors contain a singlet chiral field
$(1)_2$. As we will see in the next chapter, these three singlets
generically get   vacuum expectation values
of the order of the string scale, giving masses to the
extra colour triplets in the ${\bf 37}$ sector of the model.
In this way,  the massless spectrum of the model
coupling to the $SU(3)\times SU(2)_L\times SU(2)_R\times U(1)_{B-L}$
group  is indeed the one proposed in chapter 2. Thus, the results
for gauge coupling unification obtained there directly apply to the present
model.

\newpage
\begin{table}
\footnotesize
\renewcommand{\arraystretch}{1.25}
\begin{center}
\begin{tabular}{|c|c|c|c|c|c|}
\hline Matter fields  &  $Q_3$  & $Q_L $ & $Q_R $ & $Q_7$ & $Q_{(2,i)}$   \\
\hline\hline {{\bf 77} sector} & & & & &  \\
\hline $(3,2,1)$ & 1  & -1 & 0 & 0 & 0  \\
\hline $(\bar 3,1,2)$ & -1  & 0  & 1 & 0 & 0  \\
\hline $(1,2,2)$ & 0  & 1  & -1 & 0 & 0  \\
\hline $(4,\bar 7)'$ & 0  & 0 & 0 & -1 & 0 \\
\hline $(1,{21})'$ & 0 & 0 & 0 & 2 & 0  \\
\hline\hline {{\bf 37} sector} & & & & &  \\
\hline $(3,1,1)$ & 1 & 0 & 0 & 0 & -1 \\
\hline $(\bar 3,1,1)$ & -1 & 0 & 0 & 0 & -1 \\
\hline $(1,2,1)$ & 0 & 1 & 0 & 0 & 1 \\
\hline $(1,1,2)$ & 0 & 0 & -1 & 0 & 1 \\
\hline $(4,1)'$ & 0 & 0 & 0 & 0 & -1 \\
\hline $(1,7)'$ &  0 & 0 & 0 & 1 & 1 \\
\hline\hline ${\bf 3 3}$ sector & & & & &  \\
\hline $(1)$ & 0 & 0 & 0 & 0 & 2 \\
\hline\hline {\bf 37} bulk & & & & &  \\
\hline $(3,1,1;2)$ & 1 & 0 & 0 & 0 & 0\\
\hline $(\bar 3,1,1;2)$ & -1 & 0 & 0 & 0 & 0\\
\hline $(1,2,1;2)$ & 0 & 1 & 0 & 0 & 0 \\
\hline $(1,2,1;2)$ & 0 & -1 & 0 & 0 & 0 \\
\hline $(1,1,2;2)$ & 0 & 0 & 1 & 0 & 0 \\
\hline $(1,1,2;2)$ & 0 & 0 & -1 & 0 & 0 \\
\hline $(1,7;2)'$ & 0 & 0 & 0 & 1 & 0 \\
\hline $(1,\bar 7;2)'$ & 0 & 0 & 0 & -1 & 0 \\
\hline $(4,1;2)'$ & 0 & 0 & 0 & 0 & 0 \\
\hline \end{tabular}
\end{center} \caption{Spectrum of Left-Right model. We present the
quantum numbers  under the $U(1)^7$ groups. The first 4 $U(1)$'s come
from the 7-brane sector. The next three come from the 3-brane sector,
these we have written as a single column with the understanding that
for instance in the {\bf 37} sector, each of the three copies have
that charge under one of the three $U(1)$'s and zero under the other
two.
\label{tabps2} }
\end{table}
\newpage

\newpage
\begin{table}
\footnotesize
\renewcommand{\arraystretch}{1.25}
\begin{center}
\begin{tabular}{|c|c|c|c|c|c|c|c|}
\hline Matter fields  &  $Q_{B-L}$  & $Q_{X} $ & $Q_{A1} $ & $Q_{A2}$ &
$Q_{A3}$ & $Q_{A4}$ & $Q_{A5}$   \\
\hline\hline {{\bf 77} Sector} & & & & & & & \\
\hline $(3,2,1)$ & 1/3  & 1 & 0 & 0 & 2 & 1 & 0   \\
\hline $(\bar 3,1,2)$ & -1/3  & 1  & 0 & 0 & -1 & -2 & 0  \\
\hline $(1,2,2)$ & 0  & -2  & 0 & 0 & -1 & 1 & 0   \\
\hline $(4,\bar 7)'$ & 0  & 2/7 & 0 & 0 & -1 & 1  &  -3   \\
\hline $(1,{21})'$ & 0 & -4/7 & 0 & 0 & 2 & -2  & 6   \\
\hline\hline {{\bf 37} sector} & & & & & & & \\
\hline $(3,1,1)$ & -2/3 & -2 & (-1,1,0)& (-1,-1,2) & 1 & 1  & -1    \\
\hline $(\bar 3,1,1)$ & 2/3 & -2 & (-1,1,0) & (-1,-1,2) & -1 & -1 & -1  \\
\hline $(1,2,1)$ & -1 & 1 & (1,-1,0) & (1,1,-2) & -1 & 0 & 1    \\
\hline $(1,1,2)$ & 1 & 1 & (1,-1,0) & (1,1,-2) & 0 & 1 &   1  \\
\hline $(4,1)'$ & 0 & -2 & (-1,1,0) & (-1,-1,2) & 0 & 0 &  -1  \\
\hline $(1,7)'$ &  0 & 12/7 & (1,-1,0) & (1,1,-2) & 1 & -1 &  4  \\
\hline\hline ${\bf 3 3}$ sector & & & & & & & \\
\hline $(1)$ & 0 & 4 & (2,-2,0) & (2,2,-4) & 0 & 0 & 2   \\
\hline\hline {\bf 37} bulk & & & & & &  &\\
\hline $(3,1,1;2)$ & -2/3 & 0 & 0 & 0 & 1 & 1 & 0  \\
\hline $(\bar 3,1,1;2)$ & 2/3 & 0 & 0 & 0 & -1 & -1 & 0 \\
\hline $(1,2,1;2)$ & -1 & -1 & 0 & 0 & -1 & 0  & 0  \\
\hline $(1,2,1;2)$ & 1 & 1 & 0 & 0 & 1 & 0 & 0   \\
\hline $(1,1,2;2)$ & -1 & 1 & 0 & 0 & 0 & -1 & 0  \\
\hline $(1,1,2;2)$ & 1 & -1 & 0 & 0 & 0 & 1 & 0  \\
\hline $(1,7;2)'$ & 0 & -2/7 & 0 & 0 & 1 & -1 & 3   \\
\hline $(1,\bar 7;2)'$ & 0 & 2/7 & 0 & 0 & -1 & 1 & -3  \\
\hline $(4,1;2)'$ & 0 & 0 & 0 & 0 & 0 & 0 & 0\\
\hline \end{tabular}
\end{center} \caption{Spectrum of Left-Right model. We present the
quantum numbers  under the 4 anomaly free $U(1)$ groups and the three
anomalous $U(1)$'s. Some of the fields in the {\bf 37} sector have
several entries for $Q_Y$ and $Q_Z$, the reason being that the fields
come in three copies which differ by those charges.
\label{tabps3} }
\end{table}

\newpage

\subsection{Anomalous $U(1)$'s and Fayet-Iliopoulos terms}

The LR model contains seven Abelian $U(1)$
factors.
Four of these terms, namely $Q_3, Q_L, Q_R$ and $ Q_7$, appear in the unitary
groups of the ${\bf 77}$ sector, whereas  the other three $Q_{(2,i)}$
come from each of the ${\bf 33}_{2,i}$ sectors. As we know, some of
these $U(1)$'s  are anomalous. The spectrum
 of the model with the
corresponding $U(1)$ charges is given in Table 2.
>From there
the matrix of mixed $U(1)$-non-Abelian
 anomalies \cite{aiq} can be computed to be
 \begin{equation}
T_{IJ}^{\alpha\beta} =  \left ( \begin{array}{ccccc}
0 & 9 & -9 & 0 & 0 \\
-6 & 0 & 6 & 0 & 0   \\
6 & -6 & 0 & 0 & 0  \\
0 & 0 & 0 & 21 & -21 \\
-2 & 1 & 1 & 1 & -1 \\
\end{array}
\right )
\end{equation}
The rows correspond to the seven factors
$Q_3, Q_L, Q_R, Q_7$ and $Q_{(2,i)}$ (same structure repeats for
$i=0,1,2$). The columns correspond to the nonabelian groups:
$SU(3)$, $SU(2)_L$, $ SU(2)_R$, $ SU(7)$ and $SO(4)$ respectively.

There are two linear independent combinations of above
generators which are free of anomalies  whereas the other five have
triangle anomalies.
\footnote{This is a generic feature of this type of models with
5-branes at just one $(a,i)$ ($a$ fixed) set of fixed points.
If branes are stuck at two different $a$'s then there are ten $U(1)$'s and
three  of them are non anomalous. If there are branes at the three sets $a=0,1,2$
then four of the thirteen Abelian factors are anomaly free.}

A possible choice for the non anomalous generators is
\footnote{It is amusing that if one looks at the $Q_X$ charges
of quarks, leptons and Higgs fields, they are identical
to the ones such fields have under the
$U(1)$ contained in the branching $E_6\rightarrow SO(10)\times U(1)$.
Here, though, there is no $E_6$ nor $SO(10)$ symmetry present.}
\beqa
Q_{B-L}\ &  = & \ -\frac{2}{3}\ Q_3 - Q_L - Q_R \nonumber\\
Q_X\ & = & \ Q_R-Q_L-\frac{2}{7}Q_7+2\sum_i Q_{n_2^i}\nonumber\\
\eeqa
while anomalous ones can be chosen as
\beqa
Q_{A1}\ & = & \ Q_{n_2^0}- Q_{n_2^1} \nonumber\\
Q_{A2}\ & = & \ Q_{n_2^0}+Q_{n_2^1}-2Q_{n_2^2}\nonumber\\
Q_{A3}\ & = & \ Q_3-Q_L+Q_7\nonumber \\
Q_{A4}\ & = & \ Q_3-Q_R-Q_7\nonumber \\
Q_{A5}\ & = & \ \sum_i Q_{n_2^i} + 3 Q_7
\eeqa
 Notice that even though  $Q_{A1}$ and  $Q_{A2}$ present no mixed
$U(1)$-non-Abelian
 anomalies, they have cubic and mixed $U(1)$ anomalies.
Under these new combinations, the charges of the particles in the
spectrum are displayed in Table 3.

As usual in Type I theory \cite{sagno}, $U(1)$ anomalies are cancelled by a generalized
Green-Schwarz mechanism through the coupling to twisted close string RR fields
\cite{iru} .  Anomalous $U(1)$´s become massive \cite{pop} .
At the same time, because of supersymmetry,
 a Fayet-Iliopoulos term, associated to each of the anomalous
 groups, appears \cite{iru,iru2,pop,klein,acd}. The corresponding D-term potential is
\beq
V_r\ =\ \frac{1}{2} \left(\ \xi_r+\sum_l q_r^l|\phi_l|^2\right)^2
\label{potfi}
\eeq
where $\phi_l$ is the scalar field with charge $q_l^r$ under the
anomalous
group $U(1)_{Ar}$.
The    $\xi_r$ $r=1,\dots 5$ terms can be explicitly computed (see eq. 3.20 in Ref. \cite{aiq}) in
terms of the fields $M_{(a,i)}$, the Neveu-Schwarz partners of the RR
antisymmetric forms mentioned above.  We find
\beqa
\xi_1\ & = & \ \frac{3\sqrt{3}}{2}\ \left(\ M_{02}-M_{12}\ \right)
\nonumber\\
\xi_2\ & = & \ \frac{3\sqrt{3}}{2}\left( \ M_{02}+M_{12}-2M_{22}\ \right)
\nonumber \\
\xi_3\ & = & \frac{\sqrt{3}}{2} \sum_i \left(12 M_{0i}+
4M_{1i}+5M_{2i}\right)\nonumber \\
\xi_4\ & = & \  -\frac{\sqrt{3}}{2} \sum_i\left( 4M_{0i}+12 M_{1i}+5M_{2i}
\right) \nonumber \\
\xi_5\ & = & \  \frac{\sqrt{3}}{2} \sum_i (7M_{0i}+7M_{1i}+8M_{2i})
\eeqa
Notice that all the FI-terms $\xi _r$ are linearly independent
combinations of the $NS-NS$ twisted moduli. This means that,
unlike what usually happens in the perturbative heterotic vacua \cite{dsw} ,
there is no need to  check for D-flatness of the scalar
potentials eq.(\ref{potfi}). This is  because for any field direction
of the scalars $\phi _l$ charged under each anomalous  $U(1)_r$,
there will be vevs for the twisted moduli yielding $\xi _r$'s
compensating them. Furthermore,  this indicates, in general, a departure of
 the orbifold limit and also that the
anomalous $U(1)$'s do not remain as
effective  global symmetries, as it would have happened
 if all twisted moduli were vanishing \cite{iq} .

\newpage
\subsection{The structure of mass scales}

The unification of coupling constants in this model takes place
at a scale of order $9\times 10^{11}$ GeV which should
then be identified with the string scale $M_s$. In addition that is
also the order of magnitude of the compactification scales
$M_1$, $M_2$ of the radii of the first two complex dimensions.
This is desirable for two reasons: 1) Since the worldvolume
of 7-branes includes the first two complex dimensions,
if $M_{1,2}$ were much smaller there would be charged Kaluza-Klein
fields which might spoil gauge coupling unification; 2) Some
phenomenologically interesting
non-renormalizable Yukawa couplings involve (see next chapter)
3-branes living at different locations in the first two
compact directions. Such couplings would be considerably  suppressed
if $M_{1,2}$ were much smaller than $M_s$. On the other hand,
the compactification scale $M_3$ along the third complex plane
(which is transverse to the 7-branes worldvolume) is
unconstrained by these considerations. The Planck mass
is related to the string scale $M_s$ and the compactification
scales $M_i$ by (see e.g. \cite{imr} ) :
\beq
M_p\ =\ { {2\sqrt{2} M_s^4}\over {\lambda M_1M_2M_3} }
\ =\  {{\sqrt{2}}\over {\alpha _7}} { {M_1M_2}\over {M_3} }
\eeq
where $\alpha _7\ =\ { {\lambda M_1^2M_2^2 }\over
{2M_s^4} } $ is the unified coupling of the group
coming from 7-branes, which includes the standard
model group. Thus, for $M_{1,2}\propto M_s =9\times 10^{11}$
GeV, one can obtain the measured $M_p$ for
$M_3\propto (100)/\alpha _7 $ TeV.
In this scheme (see  fig. 5) the size of the $Y_3$ coordinate
would  be thus very large compared to $Y_{1,2}$.

The present  class of models contain anti-3-branes in the
bulk in transverse space. As depicted in fig. 5,
their worldvolume does not have
overlap with that of the ``visible world'' of 3-branes and
7-branes once the latter are located at the origin
in the third compact dimension. Anti-3-branes
are instead in  the bulk in that dimension.
The global
configuration of the model is non-supersymmetric,
since the supersymmetries preserved by branes are
broken by the anti-branes and viceversa\footnote{It is worth pointing out that even though the presence of
the anti-branes explicitly break supersymmetry, the number of massless
bosonic degrees of freedom still matches the number of massless
fermionic degrees of freedom as can be easily seen in all models of
this type, following the general spectrum of reference
\cite{aiq}.}.
Closed string states living in the bulk
of space will generically communicate supersymmetry breaking
from the anti-3-brane sector to the visible sector of
3-branes and 7-branes.
We will assume that the presence of SUSY-breaking anti-3-branes
in the bulk  constitutes a SUSY-breaking hidden sector
for this model. Since these anti-3-branes live far away in the
bulk of the (very large) third complex dimension, SUSY-breaking effects
in the 7-branes and 3-branes where the SM resides will be
Planck mass suppressed. Thus one expects SUSY-breaking soft terms
of order:
\beq
M_{soft}\ =\ \epsilon {{M_s^2}\over {M_p}}
\eeq
where the value of the fudge factor $\epsilon $
will depend on the details of how SUSY-breaking effects in the
antibranes are transmitted to the branes by the
massless closed string fields.
Since gauge coupling unification predicts $M_s=9\times 10^{11}$ GeV,
in order to get soft terms of order, say 1 TeV, we need
\footnote{This seems to suggest a one-loop transmission of
SUSY-breaking to the observable D-brane sectors, as
occurs for example in moduli dominated
\cite{spurion}   and/or  anomaly mediated \cite{anom}  scenarios.}
$\epsilon \propto 10^{-2}$.

The above assumption of a very large $Y_3$ dimensions
is a possible simple explanation for the observed
large size of $M_p$ compared to our predicted
$M_s=9\times 10^{11}$ GeV. Recently an alternative explanation
has been proposed \cite{randallsundrum} to obtain such an effect which
may occur  (in some simple models)
even if the extra dimensions are infinite.
This occurs due to the presence of  warp factors
in the space-time metric exponentially depending
on the extra dimensions. Furthermore, it has also been
argued \cite{verlinde}
that a localized set of $D3$ branes
does indeed induce a warped geometry around its location.
It would be interesting to explore whether this kind
of arguments extend to configurations like the one discussed
here which involve intersections of both 3-branes and 7-branes
\footnote{For recent studies of the Randall-Sundrum scenario
in the presence of brane intersections see \cite{intersections}. }.
An exponential warp factor depending on the
dimension $Y_3$ transverse to both 3-branes and 7-branes
could in this case be a possible alternative origin
for the $M_p$/$M_s$ hierarchy in a model like the
one studied here.

\subsection{Yukawa couplings and conservation rules}

The general structure of renormalizable couplings
in this class of orientifolds was already discussed in ref.\cite{aiq} .
Let us review the couplings involving the supersymmetric sector
for the present model, leaving their phenomenological implications
for the next section.

\bigskip
{ \it i) $\bf (77)^3$ couplings }
\bigskip

These have the form:
\beq
\phi^{77}
_i\phi^{77} _j \phi^{77} _k \ ,\ \  i\not= j\not= k\not= i \eeq
where $\phi^{77} _i$, $i=1,2,3$ are any of the
charged chiral fields in the
$\bf 77$ sector associated to the complex plane $i$. These type of
couplings give rise for example to quark Yukawa couplings, as we
discuss below. The coupling is proportional to the gauge
coupling constant for the $\bf 77$ gauge interactions $g_7$,
which is the one associated to the physical gauge fields.
Recall that the latter is related to the string scale
$M_s$ and the compactification scales $M_{1,2}$ of the first two complex planes
by:
\beq
\alpha _7\ =\ { {g_7^2}\over {4\pi} }\ =\ { {\lambda M_1^2M_2^2} \over {2M_s^4}  }
\eeq
where $\lambda $ is the Type IIB dilaton coupling. It is this $\alpha _7$ which
provides the boundary conditions for the running of the
gauge couplings of the left-right symmetric model.

\bigskip
{\it ii) $\bf (73)(73)(77)$ couplings}
\bigskip

These in principle only involve the $\bf 77$ sector
associated to the third complex plane \cite{aiq} :
\beq
\psi
^{73}_i\psi ^{73}_i \phi^{77}_3 \eeq
 where $i=0,1,2$ labels  the fixed
points  where the $3$-brane is localized.
 Notice that these couplings are diagonal in the $i$
label, i.e., there are no renormalizable couplings involving
different fixed points. These Yukawa couplings are also
proportional to the $\bf 77$ gauge coupling constant $g$.
\bigskip

{\it iii) $\bf (73)(73)(33)$ couplings}
\bigskip

In a similar manner there are superpotential couplings of the form
\beq
 \psi ^{73}_i\psi ^{73}_i \phi^{33}_{3,i} \eeq in which again
$i$ labels the fixed point. Again, only the $\bf 33$ chiral fields in the
third complex plane appear in the coupling. For example, we
already mentioned that there is
a coupling of this type between the  singlets  $(1)_2$  in the $\bf 33_i$
sectors  and the coloured triplets  in the $\bf 73_i$ sectors.
Notice however that the gauge coupling $\tilde g$ is now different, with
${\tilde {\alpha }} = \lambda /2$.

Several comments concerning the above couplings are in order.
From the string point of view, these couplings are obtained from
a disk-shaped worldsheet at which boundaries three open string
vertex operators are attached. The boundaries of the disk represent
the relevant $p$-branes, $3$-branes and $7$-branes in our case.
Thus an insertion of a vertex operator of a particle in a $\bf 73_i$
sector turns a $7$-brane boundary into a $3_i$-brane  boundary (and viceversa).
This implies that, for the disk worldsheet to make sense, $\bf 73_i$
vertex insertions (for each different $i$) have to come in pairs (see figure 6).
Thus there is  a $Z_2\times Z_2\times Z_2$ symmetry which is respected
by all disk couplings. This is obviously respected in the couplings
discussed above.

\begin{figure}
\centering
\epsfysize=10cm
\leavevmode

\epsfbox{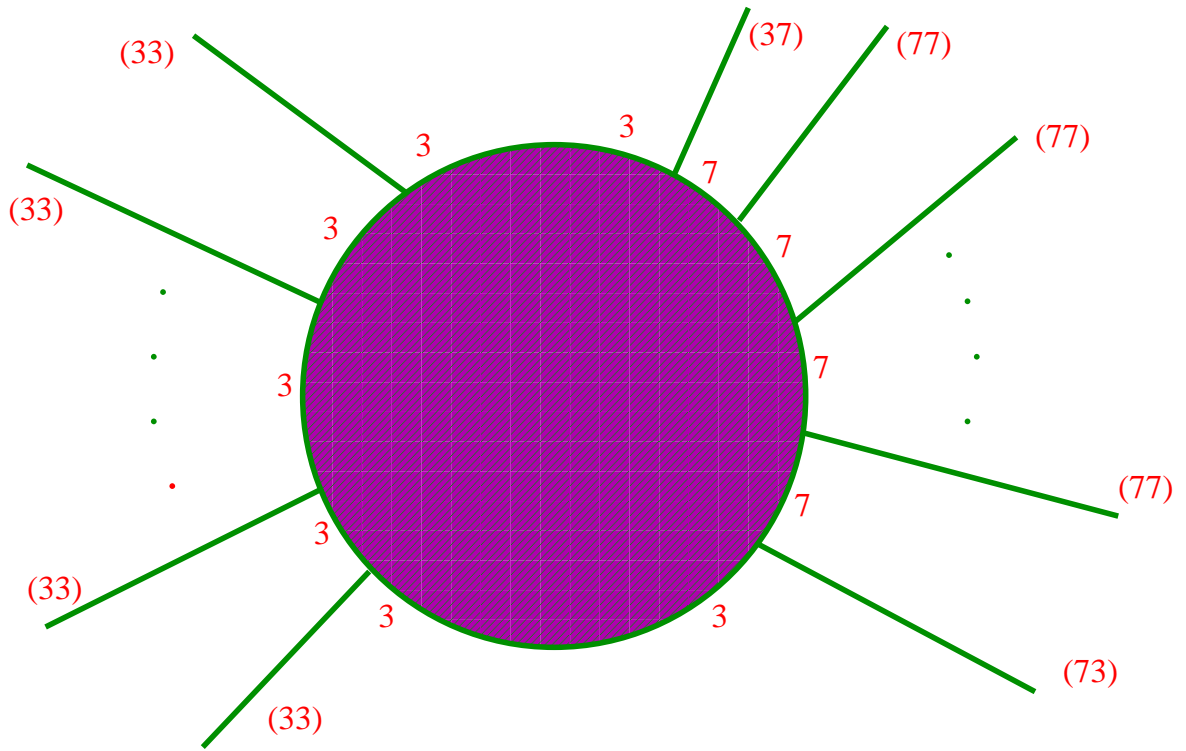}
\caption{Disk couplings of vertex operators of massless fields from ${(\bf 3
3}), ({\bf 3 7}),({\bf 7 7})$ sectors. Vertices of
$({\bf 3 7}), ({\bf 7 3})$ particles must come in pairs in order to
get a consistent D-brane boundary on the disk.}
\end{figure}

These symmetries will have an important
phenomenological role in the present model, as discussed in the next chapter.

A second question concerns the structure of the couplings $\bf (77)^3$
above. The reader familiar with heterotic orbifold constructions
will realize that the same type of couplings involving necessarily the
three different complex planes are present
for the untwisted particles  in those constructions.
In the case of heterotic orbifolds
this  antisymmetric structure may be understood in terms
of the conservation of the so called {\it H-momentum}
(see e.g., ref. \cite{fiqs,cvetic} for a discussion of these symmetries).
The right-moving vertex operators have factors proportional to the
RNS fermions which, when bosonized, can be written as
$exp(i{\bf p}.{\bf H})$ where
$\bf p$ is an $SO(10)$ (space-time)   weight.
H-momentum conservation is the statement that the overall
momentum $\bf p$ in a correlator has to vanish for a coupling
to be allowed. Now, in the Type I case something
completely analogous may be defined leading to
equivalent symmetries.

The above $Z_2$ symmetries and H-momentum conservation rules
are still valid for disk (i.e.  tree-level) amplitudes leading
to non-renormalizable couplings involving charged
open string fields. If a given
coupling involves branes living at different points
in transverse space, it will be exponentially suppressed by the
distance between those branes. Thus, for example,
couplings involving the $3_i$ branes and the $3_{bulk}$ branes
will get such a suppression. Notice however that if
the compactification scales along the first two complex planes
$M_{1,2}$ are of order the string scale $M_s$, no such a suppression
will be present. This is in fact the case considered in the previous subsection
in order to understand the hierarchy $M_p>>M_s$ : only the
third complex dimension is large and only $3$ branes distant in the
3-d complex dimension are exponentially suppressed.
This is what we will assume in the phenomenological analysis in the
next chapter.

Another point concerning non-renormalizable couplings
is  the existence of couplings violating the conservation
of {\it anomalous } $U(1)$ symmetries. Tree level couplings
renormalizable or not
should respect all non-anomalous $U(1)$ symmetries. However
they may violate anomalous $U(1)$ symmetries  since those,
 as we discussed above, are broken  by
the vevs of twisted NS-NS fields as long as we are away from the
orbifold limit (which is the case in the models discussed).
As an example of this we show in appendix B how non-renormalizable
couplings violating anomalous $U(1)$ symmetries are expected
to appear (on the basis of heterotic/Type I duality) in the
standard $Z_3$ orientifold.

\section{Phenomenology of the D-brane left-right symmetric orientifold}

In this chapter we discuss several phenomenological aspects of this model.
We will not attempt a detailed description
of all possible aspects like fermion masses or spontaneous
gauge symmetry breaking, rather we will only  discuss some possible avenues
enabling to address the gross phenomenological issues in this
model.

\bigskip
\leftline{\it i) A nearby vacuum}
\bigskip

As we mentioned in the previous chapter, apart from the
three left-right symmetric generations and Higgs fields,
this particular string model has three copies of vector-like
colour triplets form the (37) sectors. However, these states are  generically
massive. Indeed, looking at Table 3 one sees that all gauge interactions
allow for a coupling between the three singlets $1_{4}$ from the (33)
sectors to the three pairs
$(3,1,1,-2/3)_{-2}+(\bar 3,1,1,+2/3)_{-2}$, where the subindices
denote the $Q_X$ charge. These Yukawa couplings do indeed exist,
as discussed in previous chapter. Thus, if the three singlets $1_4$
get a vev of order $M_s$, the colour triplets will disappear from the low-energy
spectrum and we will be left at low energies with precisely the massless
spectrum discussed in chapter 2, leading to very good predictions for
gauge coupling unification \footnote{Notice that, on the other hand,
equivalent couplings to the Higgs doublets $(1,2,2,0)$ in the
$(77)$ sector do not exist. Thus a doublet-triplet splitting
mechanism is built-in in the symmetries of the model.}

If we give a vev $<1_4>_a\propto M_s $, $a=1,2,3$ to these (33) singlets, we
have to ensure D-flatness and F-flatness for this direction. In fact we should
not care too much about the D-terms of the anomalous $U(1)$'s
because they can be easily cancelled for appropriate values of the
blowing-up fields $M$ discussed in chapter 3. On the other hand, the
D-term corresponding to the anomaly-free $Q_X$ generator has to cancel,
which requires giving a vev to some fields with negative $Q_X$ charge.
A natural option seems to be giving vevs to the
antisymmetric $A^{ij}$  of $SU(7)$ present
in  the $(77)_3$  sector as follows
\footnote{This also turns out to give rise to the required masses
for right-handed neutrinos.} :
\beq
       A^{12}=A^{23}=A^{34}=A^{45}=A^{56}=A^{67}=A^{71}\ =\ v
\label{flat21}
\eeq
with $v\propto M_s$. This direction can be easily seen to be D-flat and F-flat.
Below $M_R$ the only $U(1)$ interaction left is now $Q_{B-L}$ since
$Q_X$ is broken by the above vevs.

\bigskip
\newpage

\leftline{\it ii) The breaking of the left-right symmetry and bulk 3-branes}

\bigskip

The chiral multiplet content of our model below $M_s$ includes
just thee generations of quarks/leptons/Higgs fields.
Some additional fields (particularly, $SU(2)_R$ doublets) are
needed if we want to break our theory down to the SM gauge group.
Probably there is more than one way to modify the model
in such a way that one has additional massless $SU(2)_R$
doublets for symmetry breaking while the good coupling unification
predictions are not spoiled
\footnote{In particular, the variant left-right symmetric model
displayed in table 1 of ref.\cite{aiq} is another possibility. That model
has additional matter but one can check that gauge coupling unification
along similar lines to those of the present model takes place.
A discussion of this variant model is presented in appendix A}.
The simplest possibility seems to add some additional 3-branes
moving in the bulk in the first two compact directions but at the
origin in the third compact direction (so that their worldvolume overlaps
with that of 7-branes). The simplest set of 3-branes that one can add in the
bulk are 6 of them (one 3-brane and their orbifold and orientifold
mirrors). They lead to a $SU(2)$ gauge group in the $\bf (33)_{bulk}$ sector and
massless chiral fields in the $\bf (73_{bulk})$  sector transforming like:
\beq
[(3,1,1, -2/3;2)\ +\ (1,2,1,-1;2)\ +\ (1,1,2,+1;2)\
+h.c.]\  +\ (1,7;2)'+h.c.\ +\ (4,1;2)' \ .
\label{bulky3}
\eeq
In addition there are chiral fields in the $\bf (33)_{bulk}$  sector
transforming
like $2(1)+(3)$ under the $SU(2)$ group on the 3-branes.
We will see later on when we discuss neutrino masses that
the $SU(2)$ group coming from this bulky 3-branes should be
broken close to the $M_s$ scale by vacuum expectation values
of the fields $(1,7;2)'+h.c.$ above.

The chiral fields in eq.(\ref{bulky3}) include $SU(2)_R$
doublets which can in principle get a vev and break the symmetry.
Thus we will assume that some  of the fields in
$(1,1,2,+1;2)+(1,1,2,-1;2)$ will get vacuum expectation values of
order $M_R\propto 1 TeV$ and break the symmetry to that of the SM.
We will briefly discuss below how that could take place due to
a radiative symmetry breaking mechanism.

One interesting point of the extra particle content provided by the
addition of these ``bulky 3-branes" is that they give a net vanishing
contribution to the
combinations $(B_1'+B_L-{{14}\over 3}B_3)$ and
$(B_L-{3\over {11}}B_1')$ which control the joining of coupling constants.
Indeed we  can easily check that extra contributions
to the $\beta $-functions are obtained:
\beq
\Delta B_L\ =\Delta B_R \ =\ \Delta B_3 \ =\ +2 \ ;\
\Delta B_1' \ =\ \Delta B_R +{1\over 4} \Delta B_L\ =\ {{22}\over 3}
\label{delbulky}
\eeq
so that unification of couplings is not modified
\footnote{This is analogous to the well known fact that complete
$SU(5)$ representations do not modify the one-loop
conditions for unification in the MSSM.}at one loop.

\leftline{\it iii) Quark and charged lepton masses}

\bigskip

In this model  renormalizable
quark Yukawa couplings of type $(77)^3$ exist with the structure:
\beq
g\ \epsilon_{ijk} (3,2,1,1/3)_i({\bar 3},1,2,-1/3)_j(1,2,2,0)_k
\label{yukquarks}
\eeq
where $i,j,k=1,2,3$ label the three complex planes
and g is the  (77) gauge   coupling constant.
With this simple structure, there would be a massless
quark generation and two degenerate generations with masses of
order $g\sqrt{\sum_i |<H_i>|^2}$. However, this structure is modified
by various effects.  To start with,
 the Kahler metric
for the $(77)$ matter fields in a model like this one needs not be diagonal.
 There are
Kahler untwisted moduli which mix the different complex planes.
Furthermore, other effects mixing different complex planes
may come from non-renormalizable D-terms like, e.g.
$<21_i21_j^*>Q_R^iQ_{R}^j*$. In addition to these,
there are mixing terms with the color triplets from the
bulky branes.
In particular
 there are renormalizable Yukawa couplings of the
form $(77)(73_{bulk})^2$:
 \beqa
& ({\bar 3},1,2,-1/3)_3\times ( 3,1,1,-2/3;2) \times  < (1,1,2,1;2)> & \\
\nonumber
&  (3,2,1,1/3)_3\times ({\bar 3},1,1,2/3;2) \times  < (1,2,1,-1;2)> &
\label{mixingq}
\eeqa
and hence the right-handed D-quarks  of the (77) sector
mix with the colour triplets from the $(73_{bulk})$ sector once
the $SU(2)_R$ doublets in that sector get a vev.
This means that generically two  physical right-handed D-quarks 
(the three of them in the variant model of the appendix) will
be  $SU(2)_R$  singlets and the other will be contained in a doublet
\footnote{This is analogous to the {\it alternate}
left-right models considered in refs.\cite{ma1,ma} .
Those models are interesting from the point of view of
supression of FCNC, as we comment below.}.
The second Yukawa coupling above will give masses
to the first two D-quarks and the couplings in (\ref{yukquarks})
will give masses to the third.

Concerning the possible Yukawa couplings for the leptons, the
following couplings in the disk are allowed by all non-anomalous
gauge symmetries:
\beq
g\ (1,2,2,0)_3 \times   (1,2,1,-1)_a \times  (1,1,2,+1)_a
\times f_a (M_{\gamma})
 \ a=1,2,3
\label{leptonyuk}
\eeq
i.e., the Higgs fields along the third complex plane in the $(77)$
sector couple diagonally to the lepton generations.
 Looking at Tables 2 and 3 we  can observe
that these couplings are indeed allowed by all non-anomalous symmetries,
including the $Q_X$ generator. However they are in principle forbidden by the
anomalous $U(1)$'s which are spontaneously broken at the string scale.
The twisted moduli fields discussed in the previous section
are charged (non-linearly) under these anomalous $U(1)$'s so one expects that
upon the insertion of coherent sets of twisted vertex $M_{\gamma}$ operators
the above couplings will be allowed, as discussed
in previous chapter and exemplified in appendix B.
 This we denote by the addition
of the factor $f_a(M_\gamma )$ in the above expression. Notice that
this factor does not necessarily mean an exponential suppression, since
the physical gauge couplings of the SM  gauge interactions are given
by the couplings on the $(77)$ sector which are proportional to
$\lambda /(M_s^4R_1^2R_2^2)$, but not to the dilaton $\lambda $ itself.
Thus $\lambda $  need not be too small a number
(see also the discussion in appendix B). The precise
size of the obtained lepton masses depends on the size of the vev
for $(1,2,2,0)_3$ and on the value of $f_a(M_\gamma )$ for each
$a$.

\bigskip

\leftline{\it iv) Neutrino masses}

\bigskip

In this model there is no right-handed $SU(2)_R$ triplet which might
give a large Majorana mass to the right-handed neutrinos.
However there are fields $N_a$ which are singlets  under the
$SU(3)\times SU(2)_L\times SU(2)_R\times U(1)_{B-L}$ group and
can combine with the right-handed neutrinos which then get a
Dirac mass of order $M_R$. Specifically, those singlets are contained
in the $(1,7')$ representations in the three $(37)$ sectors.
For the relevant couplings to appear we have to give vevs of order
the string scale to the $(1,7;2)'+(1,{\bar 7};2)$ chiral fields in the
$(73_{bulk})$ sector. In particular there is a D-flat and F-flat
direction along  $\Psi _6^2={\overline {\Psi}}_6^1=u$ and
$\Psi _7^2={\overline {\Psi}}_7^1=iu $, where
 in $\Psi _r^s$ (${\overline {\Psi}}_r^s $) the index $r$ runs over
$SU(7)$ and $s$ over the $SU(2)$. Then an effective renormalizable
Yukawa coupling is induced at low energies of the form:
\beq
(1,1,2,-1)_a \times  (1,1,2,+1;2) \times  (1,7)'_a  < (1,21')^6 \times (1,7;2)
h(M_\gamma ) >
\label{neutrinoyuk}
\eeq
where the $(1,1,2,+1;2)$ are $SU(2)_R$ doublets from the
3-branes in the bulk.
One can check that this coupling is allowed by all anomaly-free gauge
interactions of the model. As happened with the masses of charged leptons,
insertions of twisted moduli fields will be required, which we parameterize
by the factor $h(M_\gamma )$. Notice that this coupling, since it is
non-renormalizable,  is  in principle suppressed by powers of $M_s$.
It is of the general form $(73)^2(73_{bulk})^2(77)^6$ and hence involves
3-branes located at different points which will also mean exponential
suppression in the distance between the location of the 3-branes at the fixed
points and those in the bulk. Notice however that, as we discussed in the
previous chapter, we have chosen the first two complex compact directions
with sizes of order $1/M_s$ and hence there is not necessarily any extra
suppression, only the third compact complex dimension is assumed to be very
large.

Once the fields $(1,1,2,+1;2)$ get a vev breaking spontaneously
the $SU(2)_R$ symmetry, the right handed neutrinos inside the
three $(1,1,2,-1)_a$ fields will get a mass of order $M_R$ combining
with some singlets inside the $(1,7)'_a$. Notice in this connection that
generically the $SU(7)$ gauge symmetry is broken and
those fields  behave indeed like singlet partners of the right-handed
neutrinos. In this situation the left-handed neutrinos
remain massless. However there are mixing terms from analogous couplings
involving $SU(2)_L$ doublets of the form:
\beq
(1,2,1,+1)_a \times  (1,2,1,-1;2) \times  (1,7)'_a  < (1,21')^6 \times (1,7;2)
h(M_\gamma ) >
\label{neutrinoyuk2}
\eeq
Then the left handed neutrinos get induced Majorana masses of order
$m_{\nu _L}\propto m_l \times (<(1,2,1,-1;2)>/ < (1,1,2,+1;2)>)$.
The particular sizes depend on the vev of $<(1,2,1,-1;2)>$, since
$ < (1,1,2,+1;2)>$ we know is of order $M_R\propto 1$ TeV.
One thus gets neutrino masses of order:
\beq
m_{\nu}^a \ \propto m_l^a \times 
{ { <(1,2,1,-1;2)>} \over {M_R} }
\label{neutrmass}
\eeq
For $<(1,2,1,-1;2)>\propto m_l$ a  seesaw-like  formula
is obtained but the precise sizes depend on the
unknown values of the vevs of the $SU(2)_L$ doublets
$ <(1,2,1,-1;2)>$.  
Notice however that these mass contributions are flavour diagonal,
there is no mixing between different lepton families. Thus oscillations
can only take place  into some inert sterile massless neutrino
contained in the original $(1,7)'_a$ fields.
This is not a generic property of the present scenario. One can check
that in the variant model described in the appendix mixing between different
neutrino flavors can take place, since there are no $Z_2$ lepton
parities.

\bigskip

\leftline{\it v) Discrete symmetries and proton stability}

\bigskip

The couplings in this orientifold model respect a number of discrete
$Z_2$ symmetries:

i) There is a $Z_2$ symmetry associated to each of the
three $(37)$ sectors. Under it all $(73)$ particles are
odd and the rest are even. Indeed, if we consider
the  couplings of $(37)$ particles on the boundary of the disk,
they have to appear in multiplets of two
(see fig.6). Since in these sectors
live the leptons (and some singlets coming from the $(1,7')$'s
which, as we saw above behave like neutrino-like fields),
this corresponded to a discrete $Z_2$ lepton number parity.
There is one $Z_2$ symmetry for each
 of the three flavours.

ii) The flat direction considered gives vevs to the
fields $(1,7;2)'+(1,{\bar 7};2)'$ and also to some $SU(7)$ antisymmetric
fields. Thus this direction respects a $Z_2$ symmetry under which
$7$-plets and $SU(2)$ doublets (with respect to the $(33_{bulk})$ group)
are odd. Under this symmetry all quarks and leptons are
even but the $(1,7')$ fields in the $(73)$ sectors are odd.
The fields in the $(73_{bulk})$ are odd, since all are SU(2) doublets.

In fact , after breaking of the $SU(2)_R$ symmetry by the
$(1,1,2,\pm 1;2)$ fields and of the $SU(2)_L$ by the
$(1,2,2,0)$ (or, in addition,  the $(1,2,1,\pm 1;2)$ fields),
the diagonal $Z_2$ which is the combination of the original $Z_2$ and
the center of $SU(2)_R$ and $SU(2)_L$ remains still unbroken.
Thus, even after electroweak breaking a $Z_2$ symmetry remains under which

{\bf *} All quarks and leptons are odd.

{\bf *} Higgs fields breaking $SU(2)_L$ are even.

{\bf *} Singlets combining with right-handed neutrinos are odd.

{\bf *} $SU(2)_R$ and $SU(2)_L$ doublets in the $(73_{bulk})$ sector are even.

{\bf *} Colour triplets in the $(73_{bulk})$ sector are odd.

Note that this residual symmetry can be identified with the
standard R-parity of supersymmetric models.

In summary, the effective lagrangian has a residual
R-parity symmetry
\footnote{It is well known that a residual R-parity
remains in left-right symmetric models if the
$SU(2)_R\times U(1)_{B-L}$ symmetry
is broken by $SU(2)_R$ triplets $(1,1,3,-2)$.
Notice that this is {\it not} the case here and
the origin of the residual R-parity is different.}
 and in addition three lepton parities, one
per lepton flavour. It is well known that R-parity may be
considered as a discrete $Z_2$ subgroup of the B-L
symmetry. Thus combining it with the lepton parities
we thus have a $Z_2$ symmetry associated to baryon number.
This means that nucleons are stable, since a $Z_2$ baryon
parity has to be conserved under which baryons are odd and
leptons and mesons are even. Thus protons are absolutely
stable. On the other hand baryon number can be violated
in two units, since there is a $Z_2$ symmetry. This means
that in principle there can be neutron-antineutron
transitions allowed. However those transitions violate
B-L symmetry and hence are suppressed by high powers of
$(M_R/M_s)$ and the rate in this model is negligible.
Notice on the contrary that discrete symmetries
allow for the neutrino masses discussed in the
previous subsection.

\bigskip

\leftline{\it vi) Gauge symmetry  breaking and the low energy spectrum}

\bigskip

As we discussed in the previous chapter, due to the presence of
anti-3-branes in the bulk, one expects the generation of
SUSY-breaking soft terms in the effective action.
Once soft SUSY-breaking terms appear,
$ SU(2)_L\times SU(2)_R\times U(1)_{B-L}$ gauge symmetry breaking
can occur due to loop corrections. We will not perform a complete
analysis of the (quite involved) scalar potential, but will just
study what scalar fields are likely to get vevs once loop
corrections are included. As usual they will be the $SU(3)$ colour singlets
with Yukawa couplings to coloured fields. These include
the $SU(2)_L$ and $SU(2)_R$ doublets in the model, as well as the
fields in the $(33_{bulk})$ sector which are triplets $(1;3)$ under the
$SU(2)_{bulk}$ gauge group. The following Yukawa couplings
appear at the renormalizable level:
\beqa
& (3,2,1,1/3)_i \times  ({\bar 3},1,2,-1/3)_j \times  (1,2,2,0)_k &\\ \nonumber
& ({\bar 3},1,2,-1/3)_3 \times  (3,1,1,-2/3;2) \times  (1,1,2,+1;2)&\\ \nonumber
&  (3,2,1,1/3)_3 \times  ({\bar 3},1,1,2/3;2)\times  (1,2,1,-1;2) &\\ \nonumber
&    (3,1,1,-2/3;2) \times  ({\bar 3},1,1,+2/3;2) \times (1;3) & \ .
\label{yukrad}
\eeqa
The first of these couplings is the  $(77)^3$ quark Yukawa coupling
that we mentioned above. The second and third couplings are of
type $(77)_3(73_{bulk})^2$ and the fourth of type $(33_{bulk})(73_{bulk})^2$.
All these four Yukawa couplings tend to give negative mass$^2$ to the
above colour singlet scalars from one-loop diagrams in which the
colour triplets circulate in the loop.

In addition one also expects generically the presence of 
trilinear {\it  scalar}  couplings
( "A-terms") involving the colour singlet scalars.
They are proportional to the scalar  couplings
\beqa
&    (1;3) \times  (1,2,1,-1;2) \times  (1,2,1,+1;2) &\ +\ h.c.  \\ \nonumber
&    (1;3) \times  (1,1,2,+1;2) \times  (1,1,2,-1;2) &\ +\ h.c.  \\ \nonumber
&    (1,2,2,0)_3 \times  (1,2,1,-1;2) \times   (1,1,2,+1;2) & \ +\ h.c.\\
\nonumber
&    (1,2,2,0)_3 \times  (1,2,1,+1;2) \times   (1,1,2,-1;2)\ +\ h.c.  &
\label{doubyuk}
\eeqa
The first two have couplings proportional to the $(33_{bulk})$
gauge coupling constant whereas the last two are proportional to
the $(77)$ gauge coupling. The corresponding A-terms are proportional
to $M_{soft}$ and only involve the corresponding scalars.
These contributions to the scalar potential are not positive definite
and will favor all $SU(2)_R$ and $SU(2)_L$ doublets (and the scalars
in $(1,3)$) to get a non-vanishing vev at some level. We will
assume that a stable minimum of the scalar potential exists
for vevs of the order of magnitude:
\beqa
& <(1,1,2,+1;2)>\propto < (1,1,2,-1;2) > \propto < (1;3) > \propto M_R\propto
1 TeV & \\ \nonumber
&< (1,2,2,0)_i > \propto  M_Z \ << M_R &
\label{vevis}
\eeqa
so that the required hierarchy between the left and right gauge symmetries
is obtained. Notice that in principle all interactions respect an
explicit parity left$\leftrightarrow $right symmetry and it is the vacuum
which will explicitly break parity symmetry and decide who is left-handed
and who is right-handed. Whatever $SU(2)$ survives to lower energies
we will call $SU(2)_L$ by definition.

Another relevant question is what is the mass of the extra Higgs and
Higgsino fields that this model has both from the $(77)$ and
$(73_{bulk})$ sectors. This is a complicate issue which will depend
on the detailed structure of vevs.
 Looking at the first, third and fourth couplings in
eqs.(\ref{yukrad}) we see that vevs of order $M_R$ for
$(1;3)$ and $(1,1,2,\pm 1;2)$ will make massive  some of the
$SU(2)_L$ doublets in the $(73_{bulk})$ sector and also the
$(1,2,2,0)_3$ fields in the $(77)$ third complex plane.
In this situation we would  be left at low energies with the
fields $(1,2,2,0)_1$ and $(1,2,2,0)_2$ corresponding to the
first two complex planes. However, as we mentioned when we
discussed quark Yukawa couplings, there are different effects which
will generically mix the particles living in different complex
planes in $(77)$ sectors.
Thus one  also expects that
these other doublets could become massive. We will thus assume that
at a scale of order $M_R$ only one set of SM doublets remains
relatively light, so that they are available for
$SU(2)_L$ spontaneous symmetry breaking.

In addition there are the extra right-handed chiral fields
$(1,1,2,\pm 1;2)$ from the $(73_{bulk})$ sector. Some of these
where eaten in the process of $SU(2)_R$ breaking.
The remaining  may acquire a mass of order $M_R$
from the second equation in (\ref{yukrad}) , once the scalars $(1;3)$
get a vev. The same applies to the extra colour triplets
$(3,1,1,1/3;2)+h.c.$ from the $(73_{bulk})$ sector. We already mentioned
that some combination of them mixes with the right-handed quarks
from the $(77)$ sector. The orthogonal combination will get a mass
of order $M_R$  once the scalars $(1;3)$
get a vev. All in all, the spectrum below the $M_R$ would thus be
similar to that of the MSSM: three quark-lepton chiral multiplets and
one set of $H_u+H_d$ Higgs fields.

\bigskip

\leftline{\it vii)  Ramond-Ramond fields and invisible axions}

\bigskip

We already mentioned in chapter 3 that in this class of orientifold models
there are twisted Ramond-Ramond singlet scalars which couple to
${F{\tilde F}}$. As already discussed in ref.\cite{biq} , they are natural
candidates to play the role of invisible axions in a model like this.
Notice however that, in the absence of other  charged scalar vevs,
the combinations of twisted Ramond-Ramond fields coupling to the gauge
groups get in fact large masses of order the string scale $M_s$ 
by providing the longitudinal degrees of freedom of anomalous $U(1)$'s
when the latter become massive. This can be easily seen from eq.(\ref{potfi}).

Now, again from eq.(\ref{potfi}), since $\xi_r\neq 0$,
in the presence of other charged fields from the open string
sector acquiring a vev and contributing to the anomalous $U(1)$ breaking,
 there will be a linear combination of RR-field plus the charged
field,  which will be swallowed by the $U(1)$ to become massive. The
orthogonal combination will remain massless. This massless linear combination
will in general couple to the gauge fields (and in particular, to QCD)
in the standard axionic fashion with
the gauge kinetic function taking  the general form
$f_\alpha\ = \ S + s_\alpha^{(ai)}\ M_{(ai)}$ with $s_\alpha^{(ai)}$
constant computable coefficients \cite{iru2,acd}\  and
with a  decay constant of order the
string scale $M_s\propto 10^{12}$ GeV, well within astrophysical
limits.

\bigskip

\leftline{\it viii)  Experimental signatures}

\bigskip

The most obvious experimental implication of the present scheme
is the existence of extra $W_R$, $Z_0'$ gauge bosons
corresponding to the left-right symmetric gauge interactions
at a scale of order 1 TeV or below.
 The phenomenology
of $SU(3)\times SU(2)_L\times SU(2)_R\times U(1)_{B-L}$
models has been extensively studied in the past, although
most of the studies have tacitly assumed an $SO(10)$
embedding of such gauge symmetry leading to the
canonical value for the weak angle
\cite{mohap} . In addition, many
studies have concentrated on a scheme in which
$SU(2)_R$ chiral  triplets transforming like $(1,1,3,-2) + h.c.$
break the left right symmetry. At the same time these vevs
could give rise to large Majorana masses for the right-handed neutrinos,
leading to a see-saw structure for neutrino masses. This kind
of Higgs fields do not appear in the class of models that we construct,
and right-handed neutrinos are expected to become massive by
combining with other singlet chiral fields, as explained above.
Thus many  previous studies do not directly apply to
the present model.

There are a number of experimental limits on the masses
of the extra gauge bosons \cite{bkmr} .
If right-handed neutrinos are lighter than the $W_R$ mass,
the channel $W_R\rightarrow l_R\nu _R$ is open leading to clean
signatures at the Tevatron. From searches in that channel
$D0$ has set \cite{D096}
 the limit $M_{W_R}> 720$ GeV and
$CDF$ $M_{W_R}> 650$ \cite{CDF96} .
 If right-handed neutrinos are heavier
than the $W_R$, this  signature disappears and weaker
limits coming from dijet production are obtained.
$D0$ excludes the range $340<M_{W_R}<680$ GeV
\cite{D097}  whereas
$CDF$ excludes $300<M_{W_R}<420$
\cite{CDF97} . UA2 had  excluded
the energy range $100<M_{W_R}<251$ also from the dijet signature
\cite{UA2} .
There are stronger constraints on the  $W_R$ mass
from the $K_L-K_S$ mass difference but those are much more
model dependent
\cite{PDG}. On the other hand, for left-right symmetric
models with $g_L=g_R$ like this,  one can obtain limits
from precision LEP-I measurements and
low-energy neutral current  data
yielding $M_{Z_0'}>900$ GeV, implying $M_{W_R}>780$ GeV
in this class of models
\cite{cln} . In summary, the extra
gauge bosons appearing in a left-right symmetric model like this
should weight more than around 800 GeV or so
\footnote{If the right-handed D-quarks are mostly $SU(2)_R$
singlets as discussed above,  
$W_R$ production is very much supressed and direct
limits on the $W_R$ mass are much weakened \cite{ma1}. 
However that is not the case for the $Z$' and hence 
mass limits of that order are expected to still apply.}
Masses of this size or a bit higher are compatible with the gauge
coupling unification results in chapter 2. However those
coupling unification results seem to prefer not very high masses
for $W_R$ and $Z_0'$.
\footnote{ It has been recently pointed out
\cite{el}  that a small amount
of missing invisible width in $Z$ decays at LEP I as well as
atomic parity violation experiments could perhaps indicate
already the existence of some extra $Z$' with mass of order
500-1000 GeV.}Thus the extra left-right symmetric degrees
of freedom could perhaps soon be discovered.

In addition in this class of models there is a triplication of
the number of Higgs fields transforming like $(1,2,2,0)$.
Thus one also expects to find at energies of order 1 TeV,
charged and neutral Higgs and Higgsino fields. In fact these
fields are potentially dangerous.
Indeed, it is well known that in
generic models with multiple Higgs
$SU(2)_L$  doublets, the unitary transformations which diagonalize
the quark mass matrices do not necessarily diagonalize the Yukawa
interactions and FCNC can in principle appear.
This FCNC problem is generically present in left-right symmetric
models with a low $M_R$ scale like this. Thus
 to suppress
sufficiently such kind of transitions, the extra Higgs fields
have to be sufficiently heavy and/or the Yukawa couplings
will need to have some symmetries. The question of how to evade
the problem of FCNC in supersymmetric left-right models
with $M_R\propto 1$ TeV has been adressed in ref.\cite{ma,ma1} .
There it is shown that this problem can be avoided if
there are present some extra $SU(2)_R$ singlet  D-type quarks
mixing with the right-handed doublet quarks in the model
in such a way that the physical D-quarks are mostly
$SU(2)_R$ singlets.
In addition extra  $SU(2)_R$ and $SU(2)_L$ doublets with non-vanishing
B-L charge are also required \cite{ma} . 
 Interestingly enough this type of extra fields
and mixings 
are also present in the D-brane model here discussed.
It would be interesting to see whether in a D-brane type of model
a similar  mechanism
as in ref.\cite{ma}  could be made operative.

Finally, there are also extra fields from the sector
which is in charge of the breaking of the $SU(2)_R$ symmetry.
In order not to spoil gauge coupling unification we have seen that
the required $SU(2)_R$ doublets should come along with
the same number of $SU(2)_L$ doublets and coloured $SU(3)$
triplets. In the D-brane explicit constructions presented in this paper
such structure of extra particles is very natural and corresponds to the
addition of some extra 3-branes in the bulk of the first two
complex dimensions. All these extra particles are expected to have masses
also around $M_R$ or the SUSY-breaking scale. Thus new extra heavy lepton-like
and quark-like extra fermions and scalars are expected to be produced at future
accelerators corresponding to this sector.

\bigskip

In this chapter we have given an overall discussion of phenomenological
aspects of the particular orientifold model introduced in chapter 3.
 We believe  that
the particular $Z_3$ orientifold here discussed is tantalizingly close
to the general new scheme proposed in chapter 2. Analogous
models based on the same orientifold can be constructed which
will lead to somewhat different phenomenological properties. For example, one
can construct similar models by adding a second discrete Wilson line
breaking e.g. the original $SU(7)$ symmetry further. Another possibility
is to consider the variant left-right symmetric discussed in the appendix .

\section{Final comments and outlook}

During more than 15 years the minimal supersymmetric standard model
has been thoroughly studied as the best motivated extension of the
standard model. During the past 10, this study became more intense due to the
realization that gauge coupling unification works very well in the
MSSM as compared to the standard model. In this
article we have proposed an interesting alternative to the MSSM mostly
motivated by the structure of D-brane models. This alternative
scenario matches the success of gauge coupling unification of the
MSSM,
 with the interesting feature that unification works only if there are
three families of quarks, leptons and Higgs fields.
In the present scheme several physical mass scales
are unified: the string unification, susy-breaking and axion
scales are one and the same and of order $M_s=9\times 10^{11}$ GeV.
 As it  has been noted in several occasions,
\cite{benakli,biq}, this intermediate scale may have important
physical implications regarding neutrino masses, the strong CP problem,
ultra high-energy cosmic rays, non thermal dark matter candidates,
inflation,
etc.

We have also
presented a concrete D-brane model satisfying most of the general
properties
of the proposed scenario.
The existence of three quark-lepton generations
has an elegant explanation in these constructions:
there are three generations because there are three
complex compact dimensions and a $Z_3$ orbifold structure\footnote{This simple explanation,
first encountered in the heterotic models in \cite{iknq} ,  is to
be compared with that in Calabi-Yau compactifications of
perturbative or non-perturbative heterotic vacua in which the
net number of generations minus antigenerations  is related to the Hodge
numbers
of the CY. Those tend to be quite large in general and only for
very particular manifolds is small.}.
 The proton is stable
due to a combination of discrete symmetries including lepton and
R-parities and Yukawa couplings are obtained for
quarks, charged  leptons and neutrinos.
To our knowledge this is the first example where R-parity appears so
naturally in string theoretical models. This guarantees the existence of
an LSP and most of the standard searches for supersymmetry. Being a
left-right symmetric model, it shares many of the good properties that
have been realized over several years about LR models
\cite{mohapatra,mohap}. Recall, however, that
the Higgs structure we find differs from the standard treatment in the
sense that the fields responsible for breaking the LR group to the
standard model are doublets that belong to the spectrum of the
corresponding
string model instead of $SU(2)$ triplets, as often  assumed in the
literature \cite{mohap} .

One may ask if going to a left-right symmetric model
is unavoidable in this class of theories. Indeed, one can
construct explicit three generation
$Z_3$ orientifold models with the gauge
group of the standard model and some examples of this type
were presented in ref.\cite{aiq} . However gauge coupling unification
does not appear as  naturally as in the left-right symmetric
models here described.
Furthermore, the absence of the $B-L$
gauged symmetry makes difficult to find field directions with
a sufficiently stable proton. We believe that this is
probably a generic property and both coupling unification and
proton stability seem to point towards a left-right symmetric
extension of the SM above a TeV scale.

Since the LR scale, $M_R$, has to be relatively close to the TeV scale in
order
for gauge unification to work, this scenario can be experimentally
tested very soon.
In the first place,  the existence of new massive gauge fields for which the
experimental constraints are becoming very strong can be explored.
 The scenario
can also be tested
by looking at  the existence
of new Higgs particles at the TeV scale, which in principle can give
rise to flavour changing neutral currents, and of course with the
production of supersymmetric
particles.

There are several  aspects that remain to be understood in this   
scenario. First,
the Higgs scalar potential needs to be studied in order to understand
the possible patterns of gauge symmetry breaking.
This is  also important in order to address the issue of possible
FCNC transitions coming from the multi-Higgs structure. Then, the
detailed structure of soft supersymmetry breaking terms needs to be
addressed. This is important in order to have more information about
the possible experimental signals of supersymmetry. Other issues, such
as the origin of baryogenesis, may depend crucially on this knowledge.
A through analysis of the general structure of neutrino 
masses and oscilations in this kind of scenario would
also be important.

As for the explicit string model constructed,
it would be interesting to study different variations e.g. with
further symmetry breaking (Wilson lines)  or locations of 3-branes
which may lead to different phenomenological details.
Also interesting would be to look for similar left-right symmetric
models using  other constructions
like Type IIA orientifolds or non-perturbative
heterotic orbifolds \cite{afiuv} leading to similar spectra. 
In addition
 the standard issues of moduli
stabilization
\footnote{For some recent attempts to understand the stabilization
of large radii see e.g. ref.\cite{radii,aiq}. }
 and the cosmological constant
\footnote{For some recent ideas about the cosmological
constant problem in the D-brane context see e.g. refs.
\cite{verlinde,bmq} .}
 remain open.

It is highly remarkable that in this string model, the unification
scale coincides with the preferred fundamental string scale determined by
supersymmetry breaking, since a priori the two scales did not have to
be related. If this turns out to be true, we might say that nature has been
misleading us  for many years into the belief  that the unification scale
was much higher. In any case, we believe the present new scenario has
many appealing  properties and deserves to be considered as a serious
alternative to the MSSM.

\bigskip

\bigskip

\bigskip

\bigskip

\centerline{\bf Acknowledgements}

We acknowledge useful conversations with B. Allanach,
M. Klein, R. Rabad\'an and A. Uranga. This work has been partially
supported by CICYT (Spain), the European Commission (grant
ERBFMRX-CT96-0045) and PPARC.
G.A work is partially supported by APCyT grant 03-03403.

\newpage

\section{Appendices}

\bigskip

\subsection{ Appendix A}

{\it A variant left-right symmetric orientifold model}

\bigskip

In reference \cite{aiq}, we found two LR models with three families
in the study of $\IZ_3$ orientifold models with a single Wilson line.
The model discussed in the text corresponds to the first of such models.
Here, for completeness,
 we will briefly describe the phenomenological aspects of the
second
model.
Following the notation of chapter 3,
 the model can be defined by the following shift
vector
and Wilson lines:
\beqa
V_7\  & = & \frac{1}{3}\ \left(1,1,1;-1,-1;0,0;0;1,1,1,1,1,1,1,1\right)\nonumber \\
W\ & = & \frac{1}{3} \left(1,1,1;1,1;1,1;0;0,0,0,0,0,0,0,0\right)
\eeqa
 which leads to a $\bf 77$ sector group  $U(3)\times U(2)_L\times
U(2)_R\times [SO(2)\times U(8)]'$.
  We must also add four 3-branes at each of the six points
at $(a,i)$ ($a=0,1$ and $i=0,1,2$). The gauge group on each of the six
3-branes will then be $Sp(2)\times U(1)$. Similar to the previous
model, the anti-branes live in the bulk and they total 24 (equals to
the total number of 3-branes). Again, in the figure we only show 4 of
them,
all the other ones are located at the images points under the
$Z_3\times Z_2$ orientifold symmetry.

\bigskip
\begin{figure}
\centering
\epsfysize= 9 cm
\leavevmode

\epsfbox{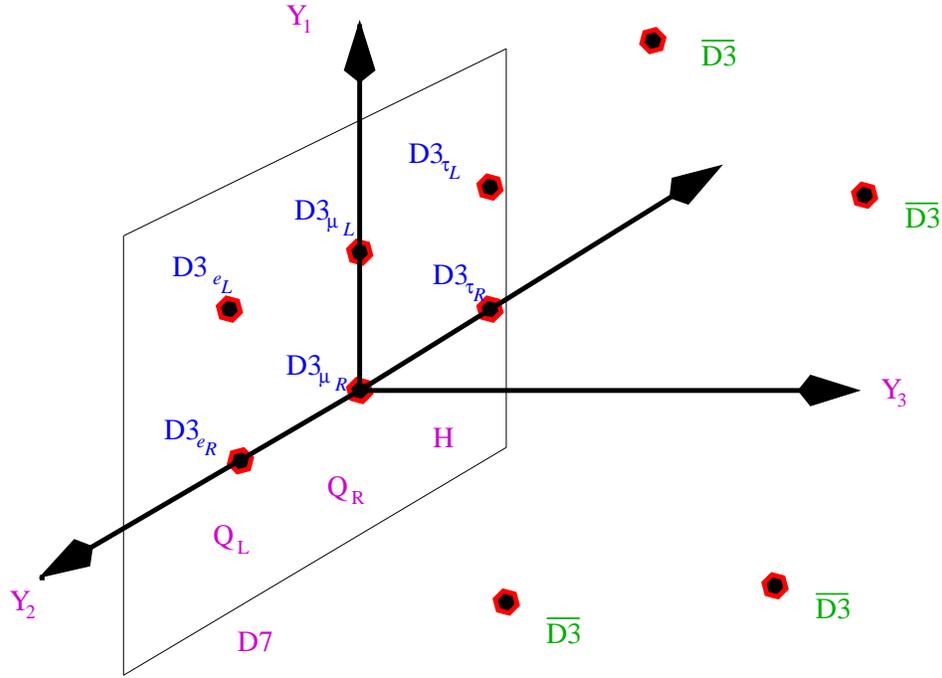}
\caption{D-brane configuration of the variant left-right model.
 The gauge group, Quarks and Higgs fields live
in the worldvolume  for the D7-brane
 whereas left-handed leptons are located at three fixed points
and right-handed leptons in other three fixed points in the complex
 $Y_1,Y_2$ dimensions. There is no need of bulk D$3_{bulk}$ branes in this
 case
for  $SU(2)_R\times U(1)_{B-L}$ breaking. The  anti-3-branes
live away in the bulk of the third ($Y_3)$ complex dimension.}
\end{figure}

Since there are no branes at the $(2,i)$ points we will have three
non-anomalous  and seven anomalous $U(1)$'s.

 We choose the linear combinations
\beqa
Q_{B-L} & = & -\frac{2}{3}Q_3+Q_L-Q_R\\
Q_{non}^L & = & {{Q_{L}}\over {2}}   +{{Q_{8}}\over {8}}\ -{2\over 3} \sum _i Q_{n_0^i}
-{1\over 3}  \sum _i Q_{n_1^i}\nonumber \\
Q^R_{non} & = & {{Q_{R}}\over {2}}  +{{Q_{8}}\over {8}}\
-{1\over 3} \sum _i Q_{n_0^i}  - {2\over 3}  \sum _i Q_{n_1^i}\nonumber\\
Q^8_A & = & Q_8 \nonumber \\
Q^r_A & = & Q_{n_a^i}
\label{sinanom}
\eeqa
with $a=0,1 $ and $i=0,1,2$.
The first 3 are non anomalous whereas the last seven are
anomalous. The spectrum with the corresponding charges under the
$U(1)$'s is shown in the table 4. We present the
quantum numbers  under the $U(1)^{10}$ groups.
 $Q_3,Q_L,Q_R $ and $Q_8, Q^A_{n_0^i}, Q^A_{n_1^i}$. The last seven can be chosen as independent anomalous combinations.
 $Q_{B-L},Q^{non}_{L}$ and  $Q^{non}_{R}$ are the  non anomalous
 charges.

\vspace*{1cm}

\begin{table}
\footnotesize
\renewcommand{\arraystretch}{1.0}
\begin{center}
\begin{tabular}{|c|c|c|c|c|c|c|c|c|c|}
\hline Matter  &  $Q_3$  & $Q_L $ & $Q_R $ & $Q_8$ &  $Q_{B-L}$ & $Q^{non}_{L}$ & $Q^{non}_{R}$& $Q^A_{n_0^i}$ &$Q^A_{n_1^i}$  \\
\hline\hline {{\bf 77} sector} & & & & & & & & & \\

\hline $(3,2,1)$ & 1  & 1 & 0 & 0 & 1/3 & 1/2 & 0 & 0 & 0  \\
\hline $(\bar 3,1,2)$ & -1  & 0  & 1 & 0 & -1/3 & 0 & 1/2 & 0 &0 \\
\hline $(1,2,2)$ & 0  & -1  & -1 & 0 & 0 &-1/2 & -1/2 & 0 & 0  \\
\hline $(2,\bar 8)'$ & 0  & 0 & 0 & -1 & 0 & -1/8 & -1/8 & 0 & 0 \\
\hline $(1,{28})'$ & 0 & 0 & 0 & 2 & 0 &1/4 & 1/4   & 0 & 0\\

\hline\hline ${\bf 37}_0$ sector  & & & & & & & & &  \\
\hline $(3,1,1;1)$ & 1 & 0 & 0 & 0 & -2/3 & -2/3 & -1/3 & 1&   0\\
\hline $(\bar 3,1,1;2)$ & -1 & 0 & 0 & 0 & 2/3 &0 &0 &0  & 0\\
\hline $(1,1,2;1)$ & 0 & 0 & 1 & 0 & -1 &  2/3& 1/2+1/3& -1 & 0 \\
\hline $(1,1,2;1)$ & 0 & 0 & -1 & 0 & 1 &  2/3& -1/2+1/3 & -1 & 0 \\
\hline $(1,2,1;1)$ & 0 & 1 & 0 & 0 & 1 & 1/2 -2/3 & -1/3 & 1&  0  \\
\hline $(1,2,1;2)$ & 0 & -1 & 0 & 0 & -1 & -1/2  & 0 & 0 &  0\\
\hline $(2,1;1)'$ &0 & 0 & 0 & 0 & 0 & 2/3 & 1/3  & -1 & 0 \\
\hline $(1,8;1)'$ & 0 & 0 & 0 & 1&0 &1/8 -2/3 &1/8-1/3& 1 & 0 \\
\hline $(1,{\bar 8};2)'$ & 0 & 0 & 0 & -1&0 &-1/8  &-1/8& 0 & 0 \\
\hline\hline {${\bf 37}_1$ sector} & & & & & & & & &  \\
\hline $(3,1,1;2)$ & 1 & 0 & 0 & 0 & -2/3 & 0 & 0 & 0&   0\\
\hline $(\bar 3,1,1;1)$ & -1 & 0 & 0 & 0 & 2/3& -1/3& -2/3& 0  & 1\\
\hline $(1,1,2;1)$ & 0 & 0 & 1 & 0 & -1 & -1/3& 1/2-2/3 & 0 & 1 \\
\hline $(1,1,2;2)$ & 0 & 0 & -1 & 0 & 1 & 0& -1/2 & 0 & 0 \\
\hline $(1,2,1;1)$ & 0 & 1 & 0 & 0 & 1 & 1/2 +1/3 & 2/3 & 0& -1  \\
\hline $(1,2,1;1)$ & 0 & -1 & 0 & 0 & -1 & -1/2+1/3 &2/3 & 0 & -1\\
\hline $(2,1;1)'$ &0 & 0 & 0 & 0 & 0 & 1/3 & 2/3  & 0 & -1 \\
\hline $(1,8;1)'$ & 0 & 0 & 0 & 1&0 &1/8 -1/3 &1/8-2/3& 0 & 1 \\
\hline $(1,{\bar 8};2)'$ & 0 & 0 & 0 & -1&0 &-1/8  &-1/8& 0 & 0 \\
\hline\hline {${\bf 33}_0$ sector} & & & & & & & & &  \\
\hline $(1,2)$ & 0 & 0 & 0 & 0 & 0 & 2/3 & 1/3 & -1&   0\\
\hline $(1,1)$ & 0 & 0 & 0 & 0 & 0 & -4/3 & -2/3 & 2&   0\\
\hline\hline {${\bf 33}_1$ sector} & & & & & & & & &  \\
\hline $(1,2)$ & 0 & 0 & 0 & 0 & 0 & 1/3 & 2/3 & 0&   -1\\
\hline $(1,1)$ & 0 & 0 & 0 & 0 & 0 & -2/3 & -4/3 & 0&   2\\
\hline \end{tabular}
\end{center} \caption{Spectrum of the variant LR model
including the charges under all $U(1)$ symmetries.  We
 have written the charges coming from the 3-brane sector as
 a single column with the understanding that
for instance in the {\bf 37} sector, each of the three copies have
that charge under one of the three $U(1)$'s and zero under the other
two.
\label{tabps2var} }
\end{table}

Mixed anomalies can be computed as we did for the model in the text and they are
exactly cancelled. The  anomaly   matrix is:
 \begin{equation}
T_{IJ}^{\alpha\beta} =  \left ( \begin{array}{cccccccc}
0 & 9 & -9 & 0 & 3 & -3 & -3  & 3\\
6 & 0 & -6 & 0 & 2& 0 &-2 & 0 \\
6 & -6 & 0 & 0 & 0& 2 & 0 &-2 \\
0 & 0 & 0 & 24 &  8 & 8& -8& -8  \\
1 & 1 &-2&  1 & 3 &0&  -3 & 0\\
1 & -2& 1 & 1 & 0 & 3 & 0 & -3
\end{array}
\right )
\end{equation}
Where the rows correspond to the 10 $U(1)$'s ordered as in
table 4. The columns correspond to the  groups:
$SU(3)$, $SU(2)_L$, $SU(2)_R$, $SU(8)'$,
 $U(1)_0^i, U(1)_1^i$ and $Sp(2)_0^i$, $Sp(2)_1^i$  respectively.

Looking at the spectrum of this model we can immediately see that
contrary to the model in the text, there are enough fields on this
model to break the group to the standard model one, {\it without}
the need to add extra branes in the bulk.
For instance, if we give a nonvanishing
vev to the $Sp(2)$ doublets in the {\bf 33} sectors we are left with a
spectrum  in the visible sector consisting only of the three families
of quarks, leptons and Higgs fields   plus  extra matter with {\it exactly}
the same quantum numbers as the `bulk' matter fields of the model in
the text (equation (4.2)), although in three copies.
\footnote{There is an ambiguity
about which doublets are assigned to be the leptons and which would be
extra matter fields. In fact we can see that the fields
$(1,2,1;2)$ in the ${\bf 37}_0$ sector and  one of the $(1,2,1;1)$ of
the
${\bf 37}_1$ sector have the same quantum numbers as the leptons. One
of the components of the $(1,2,1;2)$ field will get a mass after the
$Sp(2)$ doublet in the ${\bf 33}_0$ sector gets a vev, the remaining
component is what we will identify as the physical leptons. The other
choice
does not have the appropriate Yukawa couplings.}.
 As we have mentioned before, the quantum numbers of those
fields are such that they do not modify the analysis of the gauge
coupling unification. Therefore we conclude that, quite remarkably,
 the present model also shares the good
properties about gauge coupling unification as the model presented in
the text.
In order to cancel the D-terms generated by the vevs of the
Sp(2) doublets one has to give vevs to other doublets to
compensate. The simplest option is to take
 flat directions for the hidden sector fields
such as $(2,\bar 8)'$ of the ${\bf 37}_{0,1}$ sectors  as well as to the
$(1,28)'$ of the {\bf 77} sector. This allows in addition to generate
 nonvanishing charged lepton and neutrino masses
from couplings such  as\footnote{Yukawa couplings giving rise to quark
masses are identical to the model in the text since they only include
couplings  among fields of the {\bf 77} sector.}
\beqa
& & (1,2,2)  (1,2,1;2)^* (1,1,2;2)^* \ < (1,\bar 8;2)'(1, \bar 8;2)'
(1,28)' >
\ <(1,28)'^4>\nonumber\\
& &(1,1,2;2)^*   (1,1,2)  (1,1)_{0,i} \ < (1,2)_{0,i} (1,\bar
8;2)'(1,
\bar 8;2)'(1,28)' >
\eeqa
where the $^*$ on the $Sp(2)$ doublets stand for the component of the
doublet that remains massless after the doublet in the ${\bf 33}$
sector got a nonvanishing vev and the subindex ${a,i}$ indicates that
these are fields from the ${\bf 37}_a$ sector.
Notice that in this case the right-handed neutrinos
become massive by combining with singlets in the ${\bf 33}$ sectors.
 These
couplings are allowed by all the gauge symmetries  including the
anomalous
$U(1)$'s, therefore, contrary to the model in text,  there is no need
to
introduce insertions of twisted vertex operators in order to generate
lepton masses.

Interestingly enough, this flat direction leaves a remaining $Z_2$ symmetry
consisting in changing sign to all $SU(8)$ octets and simultaneously
changing sign to all particles in sectors ${\bf 37}$. It is easy to
see that this discrete symmetry combined with the two $Z_2$'s remaining
after breaking the LR $SU(2)$'s imply a residual discrete baryon
$Z_2$ parity
 making the proton absolutely
stable, similar to the model in the text.
Unlike that model however, lepton number is not conserved since
 couplings like
$QDL$ are permitted, although  suppressed
by powers of $(M_R/M_s)$ since they
violate $B-L$.

Combining these $SU(8)$ flat directions with those of the $Sp(2)$
doublets
does not preserve gauge coupling unification since couplings such as
\beq
(3,1,1;2)^* (\bar 3,1,1;2)^* \ < (1,\bar 8;2)(1,\bar 8;2) (1,28) >
\eeq
will in principle
give masses to the extra triplets.
Thus an alternative which does preserve gauge coupling
unification would be  combining
the $SU(8)$ flat
directions with those of singlets in the ${\bf 33}$ sectors
(not giving vevs to the Sp(2) doublets). This maintains
gauge coupling unification with the good properties for lepton masses.
In this case  giving a mass to all the extra triplets needs
 the insertion of
twisted vertex operators. In summary this
model is very similar to the one presented in the text (although probably a
bit more complicated) and may deserve
further exploration.

\newpage

\subsection{ Appendix B}

{\it Anomalous $U(1)$'s and non-renormalizable couplings in the
$Z_3$ orientifold}

\bigskip

In this appendix we will argue that non-renormalizable
couplings violating anomalous $U(1)$ charges are in
general expected in orientifolds similar to the ones
considered in the present article. More specifically
heterotic/Type I duality seems to indicate that this is the case.
Consider in particular the standard $D=4$, $N=1$, $Z_3$ orientifold
first constructed in ref.\cite{ang}. The underlying
orbifold is exactly the same than the one considered in the
present paper, the only difference being the absence of
anti-branes and   Wilson lines. The
model has 32 9-branes and a gauge group
$SU(12)\times SO(8)\times U(1)_A$. In the open string $(99)$
sector there are charged fields transforming like:
\beq
3({\overline {66}},1)_{+2}\ +\ 3(12,8)_{-1}
\eeq
where the subindex denotes the $U(1)_A$ charge.
It is easy to check that the $U(1)_A$ interaction
is anomalous.
In addition there are 27 chiral singlets $(1,1)_0$ coming
from the twisted  closed string sector. The latter are singlets under
 $U(1)_A$ but transform non-linearly under that anomalous symmetry,
and it is this transformation which cancels the $U(1)_A$ anomalies
by means of a generalized Green-Schwarz mechanism.
At the same time the $U(1)_A$ becomes massive.
Now, $SU(12)$ invariance allows for a non-renormalizable
superpotential coupling of the type $({\overline {66}},1)_{+2}^6$
by contraction with the 12-index antisymmetric tensor.
However, such a coupling would violate $U(1)_A$ conservation in 12 units.
Since this $U(1)_A$ symmetry is broken one may suspect that
such non-renormalizable coupling may however exist if
one moves away from the orbifold limit, i.e., if
the twisted NS-NS singlet fields get a vev.

Heterotic/Type I duality seems to indicate that is the case.
The above orientifold has a heterotic dual
\cite{ang,kak,afiv} which is a $Z_3$ orbifold compactification
of the $SO(32)$ heterotic string. This model has the same gauge group
and the charged particles in the untwisted sector are identical to those
in the $(99)$ sector of the Type I model. There are also
27 twisted chiral fields transforming like $(1,1)_{-4}$ under
the gauge group, which should be identified as duals of the
27 twisted fields of the Type I dual. Notice however that, unlike
their Type I counterparts,  the heterotic singlets are charged under
the anomalous $U(1)_A$. Now, one can convince oneself that
in this heterotic $Z_3$ orbifold the following
non-renormalizable couplings do in general exist:
\beq
(1,1)_{-4}^3\times ({\overline {66}},1)_{+2}^6
\eeq
This coupling preserves the $U(1)_A$ symmetry.
One  can also check that this coupling preserves the
necessary selection rules in order to be present.
In particular, it respects the $Z_3$ point group symmetry
(the twisted fields appear to the third power).
Also the H-momentum conservation is obeyed. To check this
it is enough to consider the vertex operators
associated to the $({\overline {66}},1)_{+2}$ written in the
" 0-picture". In addition, as pointed out in ref.\cite{kak} ,
the FI-term present in this heterotic model forces the singlets
$(1,1)_{-4}$ to get non-vanishing vevs. Thus away from the orbifold
point the heterotic model will present
$({\overline {66}},1)_{+2}^6$ couplings,
effectively violating the anomalous $U(1)_A$ symmetry.
This strongly suggests that similar couplings will also
be present in the dual Type I model as long as one stays
away from the orbifold point, i.e., as long as the
twisted $NS-NS$ closed string singlets have non-vanishing
vevs. This is indeed the case considered in the models in the present
article.


\end{document}